\documentclass[fleqn,usenatbib]{mnras}

\usepackage{newtxtext,newtxmath}

\usepackage[T1]{fontenc}
\usepackage{ae,aecompl}

\usepackage{graphicx}	
\usepackage{amsmath}	
\usepackage{amssymb}	

\def\SPSB#1#2{\rlap{\textsuperscript{{#1}}}\SB{#2}}

\def\SB#1{\textsubscript{{#1}}}

\title[Studying Quasar Absorber Host Galaxy Properties Using Image Stacking Technique]{Studying Quasar Absorber Host Galaxy Properties Using Image Stacking Technique}

\author[Bill Zhu, Yinan Zhao, Jian Ge, Jingzhe Ma]{
Bill Zhu,$^{1, 2}$\thanks{E-mail: duduxiong10@gmail.com (BZ)}
Yinan Zhao,$^{1}$
Jian Ge,$^{1}$
Jingzhe Ma$^{3}$
\\
$^{1}$Department of Astronomy, University of Florida, 211 Bryant Space Science Center, Gainesville, FL 32611, USA \\
$^{2}$Lynbrook High School, 1280 Johnson Ave, San Jose, CA 95129, USA\\
$^{3}$Department of Physics \& Astronomy, University of California, Irvine, 2156 Frederick Reines Hall, Irvine, CA 95129, USA
}

\date{Accepted XXX. Received YYY; in original form ZZZ}

\pubyear{2018}

\begin{document}
\label{firstpage}
\pagerange{\pageref{firstpage}--\pageref{lastpage}}
\maketitle

\begin{abstract}
Studying the stellar mass, age, luminosity, star-formation rate, and impact parameter of quasar absorber host galaxies can aid in the understanding of galaxy formation and evolution as well as in testing their models. We derive the Spectral Energy Distribution (SED) and impact parameter limits of low redshift ($z_{abs} = 0.37 - 0.55$) Mg II absorbers and of higher redshift ($z_{abs} = 1.0 - 2.5$) 2175 \AA\ dust absorbers (2DAs). We use an imaging stacking technique to statistically boost the signal-to-noise ratio (SNR) to increase detection of the absorber host galaxies. The point spread function of the background quasar is modeled with Principal Component Analysis (PCA). This method efficiently reduces the uncertainty of traditional PSF modeling. Our SED for Mg II absorbers indicates that low redshift Mg II absorber host galaxies are likely star-forming galaxies transitioning into red quiescent galaxies, with a low star formation rate of 2.2 $M_\odot$ $yr^{-1}$. From the stacked images and simulations, we show that the average impact parameter of 2DAs is > 5 times smaller than that of Mg II absorbers, at < 7 kpc instead of Mg II absorbers' 48 kpc, indicating that 2DAs are likely associated with disk components of high redshift galaxies. This means that 2DAs are likely good probes to study precursors to the Milky Way. 

\end{abstract}




\section{Introduction}

Galactic evolution is a cornerstone of astronomy that is being widely and intensely studied all over the world. One of the most powerful ways to gain insight into the properties of high-redshift galaxies is by tracking metal absorption lines in interstellar clouds. Quasars, powered by the accretion of materials in supermassive black holes, are the brightest objects in the early universe. They are able to probe metal absorption lines in intervening galaxies between the quasars and Earth, making quasar absorption line systems one of the most powerful tools in studying the gas content in the early universe to high levels of sensitivity that do not depend on redshift. 

	Bahcall \& Spitzer (1969) suggested that metal absorption lines seen in quasar spectra are induced by large gas halos surrounding host galaxies. These gas halos can extend up to 100 kpc away. Thus, tracking quasar absorption lines to study these gas halos has been an active field for the past few decades. Previous studies that investigated host galaxies of quasar absorbers have used deep imaging while the gas content has been extensively studied with spectroscopic follow-ups. However, such studies are extremely restrictive, due to small sample sizes and high expenses in performing spectroscopic follow-ups of faint objects. Thus, most studies in the late 1990s and early 2000s were limited to a few dozen case studies and focused on individual attributes of varying host galaxies. Furthermore, the wide distribution of redshifts and magnitudes of the sample result in the inability to determine strong statistical properties of the absorbers and their host galaxies. Overarching trends across different ranges of redshifts and absorption strengths could not be found. These trends are necessary for categorizing the major components found in all absorbers, such as average luminosity-weighted impact parameters for investigating location and geometry of absorbers, and average spectral energy distribution for studying stellar population, mass, and star formation history. 

	On the other hand, ground-based telescope imaging is widely available but significantly noisier, making detection of absorber host galaxies in individual frames impossible. One method of boosting the signal-to-noise ratio (SNR), and hence sensitivity, of ground-based imaging is by stacking many frames of imaging data together, relying on statistical properties of large datasets to preserve the signals of each individual frame while decreasing the overall noise level. By the central limit theorem of statistics, the noise distribution in the mean image is tighter than the original images by a factor of $\sqrt{N}$, where N is the number of frames stacked. Stacking images brings out signals that were previously undetectable in individual frames since the background noise level is greatly reduced. This not only mitigates the challenges associated with expensive, deep field, high contrast imaging follow-ups, but also offers a statistical method to study host galaxy properties as a galaxy population. Correlations between different attributes, such as surface brightness, galaxy age, mass and star formation history, can then be determined. 

	Stacking approaches and studies have provided valuable results in a number of areas. For example, Bartelmann \& White (2003) demonstrated that stacking of ROSATA ll-Sky Survey X-ray images of high-redshift clusters detected in the Sloan Digital Sky Survey (SDSS) can be used to derive their mean X-ray properties. In the context of galaxies, Hogg et al. (1997) constrained the IR signal from faint galaxies using stacked Keck data. Similarly, Brandt et al. (2001) measured the mean X-ray flux of Lyman break galaxies, Zibetti et al. (2004) characterized the very low surface brightness ``diffuse" light in galaxy halos, and White et al. (2007) constrained the radio properties of SDSS quasars down to the nanojansky level. In spectroscopic studies, stacking techniques have been extensively used to look for weak signals. Composite spectra of SDSS quasars were used to detect weak absorption lines, as well as dust reddening effects that are well below the noise level in individual spectra (Nestor et al. 2003; M\'{e}nard et al. 2005; York et al. 2006). Clearly, the stacking approach is very advantageous in investigating faint sources (Zibetti et al. 2007).

	Zibetti et al. (2005a, b, 2007) used such a stacking method for studying large samples of Mg II absorption quasars. The Mg II absorption feature was chosen as their interest of study because of dominant ions in H gas. Mg II possesses a resonance transition at the doublet (2796.35, 2803.53\AA). Zibetti et al. (2007) extracted thousands of Mg II absorber quasar cutouts from the Sloan Digital Sky Survey's Data Release 4 (DR4). They binned the Mg II absorbers by the redshift of the absorbers and by the rest equivalent width (REW) of the 2796 \AA\ absorption line. They then centered the absorbers to their calculated centroids by way of interpolation in order to make the images super-imposable, subtracted the quasar's point spread function (PSF) to diminish the quasar's light contributions, and finally rescaled the intensity of the absorbers to achieve normality and consistency. Zibetti et al. (2007) corrected the images for galactic extinction and mean-stacked the images to reveal one final mean frame. Their results indicate that there is no significant redshift dependence for both impact parameter and rest-frame colors for redshifts up to $z_{abs} = 1$. They also showed that stronger absorption systems display the colors of blue star-forming galaxies while weaker absorption systems mostly originate from red passive galaxies. Finally, Zibetti et al. (2007) demonstrated the stacking technique's usefulness in detecting the light of QSO hosts and their environments. 
  
	Since 2007, there have been numerous advances in cataloging quasar absorber host galaxies. Bouche et al. (2007) discovered that there may exist a correlation between star formation rate and the equivalent width of the Mg II absorption doublet, indicating that gas content can possibly be used as a predictor for starburst phenomena. Chen et al. (2010) used 70 low impact parameter Mg II absorbers from the Magellan Echellette spectrograph to determine that increased impact parameters of the absorbers result in decreased absorber strength. Kacprzak et al. (2012) used 88 spectroscopically confirmed Mg II absorbers from the Hubble Space Telescope and SDSS to demonstrate that there exists an azimuthal bimodal distribution of absorbers, with blue star-forming galaxies driving the bimodality. They indicated that halo gas exists more commonly around the projected galaxy's major and minor axis. In addition, the bimodality is generated by the accretion of gas along the galaxy's major axis and outflowed along the galaxy's minor axis. This is consistent with galaxy evolution scenarios where star formation galaxies accrete gas. In these same scenarios, red galaxies typically have smaller star formation rates due to lower gas reservoirs. Nielsen et al. (2013) cataloged 182 spectroscopically identified, high redshift ($0.7 - 1.1$) Mg II absorbers, indicating that low stellar mass galaxies tend to be bluer and hence possess higher star formation rates. However, they also indicated that the Mg II absorption is preferentially weaker in such systems. As shown, all of these studies required a lot of dedicated spectroscopic data acquired from telescopes and used a small sample size.
  
	A recently discovered type of absorber that is of great interest is the quasar 2175 \AA\ dust absorbers (2DAs, Wang et al. 2004, Ma et al. 2017, 2018a). These dust absorbers, displaying strong broad 2175 \AA\ absorption features and dust extinction (e.g., Wang et al. 2004; Zhou et al. 2010; Jiang et al. 2010a, b, Jiang et al. 2011; Wang et al. 2012; Zhang et al. 2015; Pan et al. 2017; Ma et al. 2015; 2017), closely resemble the Milky Way (MW). The similarity between the extinction curve of these absorbers and that of the MW means that studying these systems might provide clues on the evolution of the MW. 2DAs are also a subgroup of Mg II absorbers, representing about 1 per cent of the total Mg II absorber population detected (Zhao et al. 2018, in prep). While extensive studies on Mg II absorbers have been conducted, as shown above, very few studies on 2DAs and their host galaxies have been made. Ma et al. (2017, 2018a) performed correlation analysis between metallicity, velocity width, redshift, depletion level, and other quantities to compare 2DAs with other absorption lines. They concluded that the 2DA host galaxies contain high metallicity, high depletion, are generally massive, and are chemically enriched. 2DAs are also more massive than typical Damped-Lyman-alpha (DLA) and sub-DLA galaxies, showing greater maturation and age. The median estimated stellar mass of 2DA host galaxies is $2 \times 10^{10}$ $M_\odot$ (Ma et al. 2018a).
  
	Thus, studying 2DAs will also give insight into the dust attenuation in high redshift galaxies, which is consequential in the theory of galactic evolution. Early results (Ma et al. 2017, 2018a, b) show that 2DAs are likely matured and massive. They also report that the impact parameter for one 2DA system at $z_{abs} = 2.12$ is only 5.5 kpc, or 0.65 arcsec away from the quasar. However, this is only an isolated case as acquiring HST time is extremely difficult and it is hard to study a large number of absorber host galaxies using HST. For a sample of 436 2DAs identified by us in the SDSS-III DR 12 imaging data (Zhao et al. 2018, in prep), HST imaging for each individual absorber system is cost inhibitive. Therefore, only stacking is a viable option that will result in the sensitivity necessary for detection and for studying the overarching statistical properties of all 2DA systems.
  
	Our research is to develop an automatic and fast imaging stacking and subtraction method using data from ground-based telescopes to study the statistical properties of host galaxies of Mg II absorbers and 2DAs i.e. their average impact parameter and SEDs. The impact parameter will possibly reveal the geometry of host galaxies and will also help identify which components of host galaxies are observed through quasar absorption. In order to study the host galaxy properties and impact parameter distributions of the newly discovered and extremely faint 2DAs, we create a stacking technique that is specialized in reducing noise levels as effectively as possible. We use better and more accurate stellar profile approximation methods to decrease the noise level and boost detection. We validate our technique by comparing our results with Zibetti et al.'s (2007). Because of the large datasets, we create an automated package for ease of processing. This package can also be used to perform stacking analysis on other types of absorbers, such as Ca II, CIV or Fe II absorption systems. Our goal is to identify aspects of galaxy evolution, discover any new properties/correlations, and confirm previous studies' results by probing the statistical properties of Mg II absorbers and 2DAs. Because the 2DAs are largely precluded in the imaging data due to the high redshift range ($z_{abs} = 1.0 - 2.5$), our improvements to noise-reduction in the stacking procedure will aid in revealing properties of 2DAs. Hopefully, this will result in new findings about the MW in the future.
  
	In Section 2 of this paper, we present our methodology of collecting imaging data and selecting both Mg II absorbers, 2DAs, and their corresponding reference QSOs. In Section 3 we present the image processing techniques in terms of subtracting the sky background, masking unwanted sources, calibrating image intensities, etc. In Section 4 we calculate surface brightness (SB) profiles and classify the derived SED of Mg II absorber host galaxies. We also perform simulations of the 2DA host galaxies' average impact parameter. In Section 5 we discuss our findings. In Section 6 we compare our technique to previous studies.

\section{Data}

All data is retrieved from the Sloan Digital Sky Survey Data Releases 7 (DR7, Abazajian et al. 2009) and 12 (DR12, Alam et al. 2015). Both SDSS DR7 and DR12 provide large, detailed quasar catalogs (DR7QSO, Schneider et al. 2010; DR12QSO, P\^{a}ris et al. 2015) that contain the RA and DEC values of the quasar, the run, rerun, frame, and camera numbers, magnitudes in $g$, $r$, and $i$ bands, redshift, etc. The selection criteria of quasars is detailed in Ross et al. (2012). We retrieved all the quasar field images from the SDSS DR7 and DR12 databases in the five available color bands: $u$, $g$, $r$, $i$, $z$. We also retrieved the corresponding fpObj fits binary table file that contains a detailed list of the physical coordinates in the field image, PSF counts, object type, flags, etc. The quasar's ID number in the fpObj file is given in the DR7 QSO catalog for the Mg II absorbers and in the DR12 QSO catalog for 2DAs, so obtaining the physical coordinates of the quasar is accomplished efficiently. 

For the Mg II absorbers, We use DR7 ``corrected'' fpC images, which are flat-field corrected but retain the original background signals. For the 2DA systems, we use DR12 ``frames'', which already have an accurate sky-background subtracted. Section 2.1 describes the selection of Mg II absorbers. Section 2.2 describes the selection of 2DAs. Section 2.3 details the selection of reference QSOs for both Mg II absorbers and 2DAs.

\subsection{Selecting MG II Absorption Quasars}

In order to validate the effectiveness and accuracy of our procedure, we first emulate the Zibetti et al. (2007) study's results. We used the QSO-based Mg II absorber catalog in Zhu et al. (2013) It contains > 880 Mg II absorbers in the redshift range of $z_{abs} = 0.37 - 0.55$, as shown in Figure 1. The Mg II absorbers were systematically found by comparing Right Ascension (RA) and Declination (DEC). All images used are the fpC imaging data files given by SDSS DR7, which are bias-corrected and flat-fielded but preserve the observed background levels measured in counts.

\subsection{Selecting 2175 \AA\ Dust Absorber Quasars}

The 2DAs are characterized by dust extinction. A catalog of 436 absorbers were selected from SDSS DR12 by Zhao et al. (2018, in prep). These absorbers have higher redshift than the Mg II absorbers and span a much greater redshift range ($z_{abs} = 1.0 - 2.5$). The apparent magnitude distribution of QSOs both with Mg II absorbers and 2DAs are approximately normally distributed, with 2DA quasars being, on average, 0.5 magnitudes dimmer.

\begin{figure*}
    \includegraphics[width=\textwidth]{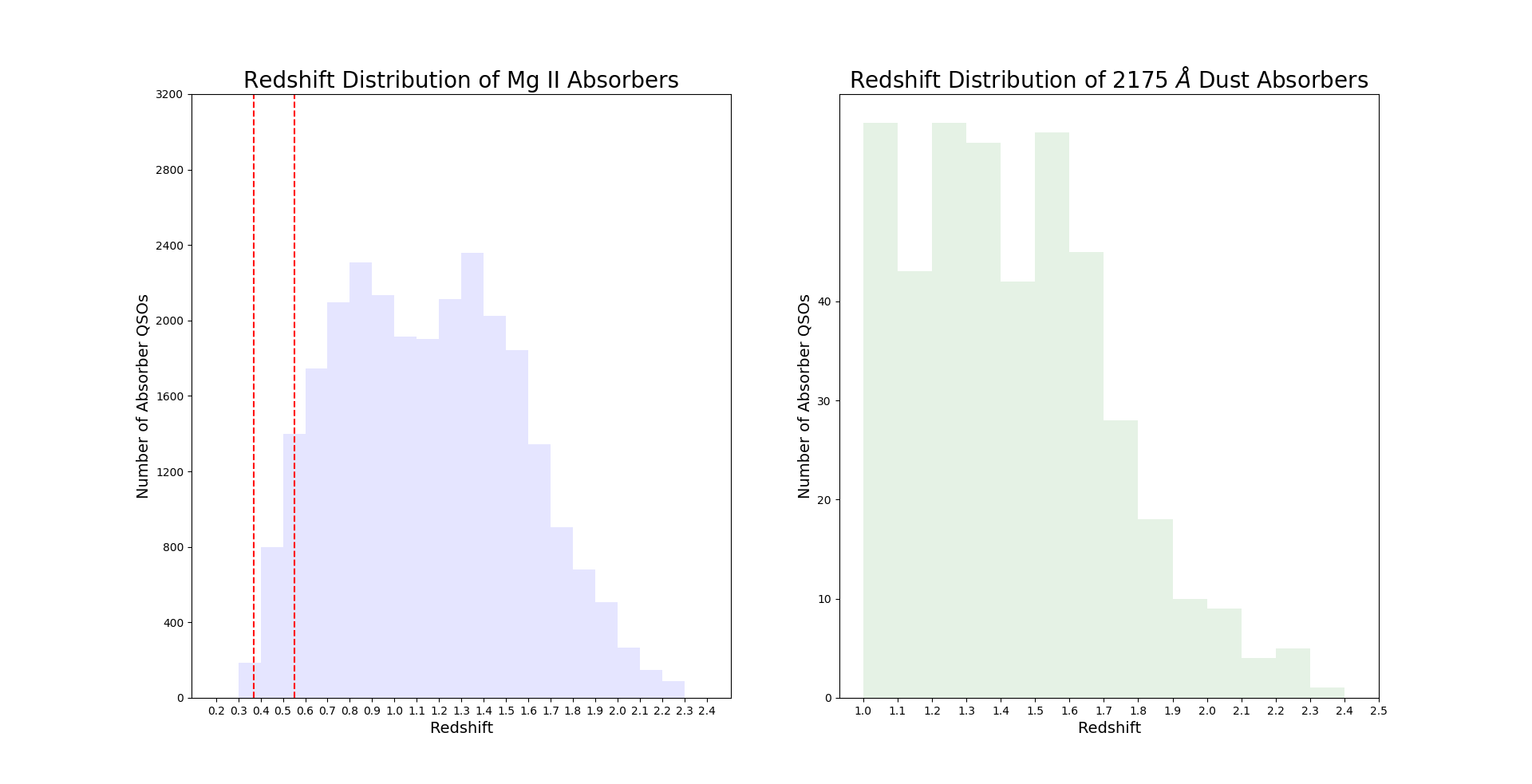}
  \caption{Above are the distributions of the redshifts of the two classes of absorbers. We will only be focusing on $z_{abs} = 0.37 - 0.55$ redshift Mg II absorbers, as indicated by the vertical dashed red lines marking the range. We will be using all the 2DAs at $z_{abs} = 1.0 - 2.5$. }
  \label{fig:Figure 1.}
\end{figure*}

\subsection{Selecting Reference Quasars}

Like any scientific study, a control sample is necessary to assert that our results are valid. Thus, a large sample of quasars that lack the Mg II absorption lines is required to prove that any results gained from quasars with Mg II absorption host galaxies are significant and unattainable from non-absorbing quasars. For each quasar with the Mg II absorption spectrum, we selected four reference quasars. These quasars must match their absorption counterpart closely. The criterion for choosing reference quasars are as follows:

\begin{itemize}
\item\ Difference in redshift must be less than 0.1
\item\ Difference in magnitude must be less than 0.5 in $ugriz$ bands
\end{itemize}

These two boundaries ensure that the reference quasars resemble the absorption quasars as much as possible so that new findings are strongly associated with the host galaxies. The quasars are then binned by the equivalent width of the 2796 \AA\ absorption line and by the redshift of the absorbers. The bins chosen were consistent with those of previous studies: 

\begin{itemize}
\item\ Redshift $z_{abs}$: $0.37 - 0.55$
\item\ Equivalent Width: $W_0 > 0.8$ \AA\
\end{itemize}

The reference quasars for both the Mg II absorbers and 2DAs are selected using the above criteria, with the exception that Mg II reference quasars are chosen from the DR7 catalog; 2DAs are chosen from the DR12 catalog. The two QSO Catalogs share many of the same attributes, some of which are listed above. For now, we only report the stacking of all Mg II absorbers in the $0.37 - 0.55$ redshift bin, in other words all Mg II absorbers as cataloged in Zhu et al.'s (2013) DR7 Mg II absorber catalog. In total, there are approximately 880 Mg II absorber QSOs, 3500 Mg II reference QSOs, 436 2DAs, and 1580 2DA reference QSOs. The exact number differs slightly among color bands, with slightly lower numbers in the $z$-band due to lower SNR and insufficient numbers of PSF stars to choose from for training the PSF.

\section{Method / Procedure}

The stacking approach capitalizes on the fact that the galaxies linked to an absorbing system statistically produce an excess of surface brightness (SB) around the absorbed QSOs with respect to unabsorbed ones. Such an SB excess can be measured to obtain both global photometric quantities for the absorbing galaxies in different bands and the spatial distribution of the light of the absorbing galaxies, from which the impact parameter distribution of absorbing gas clouds can be derived. In this section we describe the techniques that allow us to optimally integrate and measure the flux distribution of exclusively all galaxies that cross-correlate with the presence of an absorber. By ``optimally'' we mean that any source of noise is minimized. There are three main sources of noise: (1) the intrinsic photon noise, which is fixed by the number of stacked images, (2) the signal produced by field objects such as stars and galaxies which are not correlated with the absorbers, and (3) the light from the QSO itself. While the intrinsic photon noise appears to be sufficiently low for a few hundred SDSS images, the signal from background sources and the brightness mismatch that is allowed between absorbed and reference QSOs, although small, produce a noise that is orders of magnitude larger than the signal to be detected. These two sources of noise must be drastically reduced by applying accurate masking and QSO PSF subtraction algorithms on each image. The stacking and subsequent analysis is conducted simultaneously in the four bands for which most of the flux is expected, i.e., $g$, $r$, $i$, and $z$. (Zibetti et al. 2007)

Section 3.1 is devoted to the splicing of the field images to retrieve the quasars. Section 3.2 details the masking algorithm we employed. Section 3.3 explains our PSF subtraction algorithm. Section 3.4 describes the background subtraction accuracy, photometric calibration, and stacking methods used.

\subsection{Extracting Quasar Cutouts}


The quasar's location is first identified by extracting the required object ID number located under the quasar's index in the catalog. For DR7, the specific ID of the QSO, as identified in the fpObj binary table, is given directly in the DR7 QSO Catalog. For DR12, a binary decomposition method is performed to extract the ID number. After identifying the quasar, a $100 \times 100$ pixel$^2$ region centered on the quasar is cut out. 2D interpolation is then performed to center the quasar at its centroid. The background is not subtracted in this procedure, but rather when performing the PSF subtraction. The same background is calculated for both the PSF fitting and the quasar from the 3 $\sigma$ clipping median value of the image data, ensuring consistency. For the DR12 data, the background is already subtracted accurately.

\subsection{Background Source Masking Algorithm}

Zibetti et al. (2007) implemented a masking algorithm that used flux-limited masks to mask all stars with apparent magnitude < 21.0 and all sources deemed improbable to be an absorber (Zibetti et al. 2007). The circular aperture for masking is created with the median isophotal semi-major radius of the detected source which is calculated at 25 mag arcsec$^{-2}$. The actual masking radii is slightly enlarged to account for irregularities. However, as noted in DR12, the algorithm used to calculate the isophotal semi-major radius is inaccurate. In addition, circular radii are not effective in masking highly irregularly shaped sources and may unnecessarily alter pixels that do not have to be masked. Thus, we utilized a masking technique that used a flood-fill algorithm to mask non-circular objects. Once a source is identified in the quasar cut out frame by way of calculating distances between other sources and the quasar, the source is masked out spreading from the centroid of the source, recursively checking every adjacent pixel to see if the counts are more than 1 $\sigma$ above the $\sigma$ clipped mean background level of the whole image. Extended sources such as oversaturated stars are more effectively and systematically suppressed this way. In order to better constrain the masking to ensure that the quasar itself isn't accidentally masked, (due to close sources that extend a 1 $\sigma$ ``band" to the QSO) the QSO's light is first taken out by using SDSS's ``postage-stamp" images found in the fpAtlas files. This postage-stamp image is meant to be deblended from all other sources so the QSO is optimally cut out before masking takes place.

\begin{figure*}
	\includegraphics[width=\textwidth]{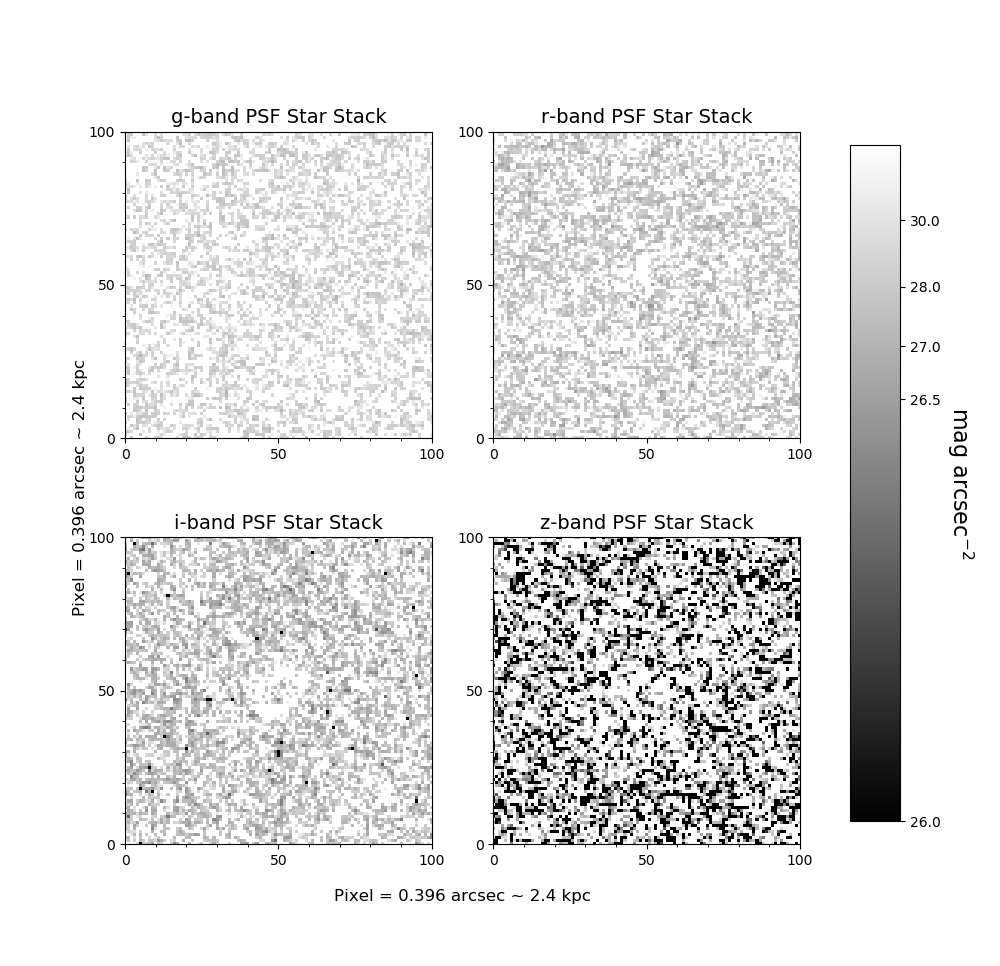}
  \caption{The PCA PSF subtracted stars stacked frames in $griz$ bands are shown here. As demonstrated, there is no significant residue in any of the bands, indicating that the PSF subtraction is good.}
  \label{fig:Figure 2.}
\end{figure*}

The above masking scheme is used for all sources identified as stars; for those identified as galaxies, an additional step is needed. In the redshift range that we are studying, $0.37 - 0.55$, some Mg II absorber host galaxies are visible in the field images, and are subsequently cataloged in the fpObj files. In order to differentiate between galaxies that contain the Mg II absorption line and those that don't, we adopt the galaxy fluxes derived for an unobscured, metal-poor stellar population produced in a 100 Myr long burst, observed right at the end of the burst itself, with rest frame $g$-band absolute magnitude $M_{g,thresh} = -22.4$. This SED is computed from the Charlot \& Bruzual (2007) SED models which updated their 2003 models. Previous research has indicated that adopting a higher threshold of $-23.9$ absolute magnitudes for computing the SED results in no systematic effect, although the number of bright interlopers significantly increases (Zibetti et al. 2007). Thus, the threshold computed with $M_{g,thresh} = -22.4$ is satisfactory. To ensure the highest consistency with Zibetti et al.'s (2007) study, we choose the same flux-limited mask parameters. If a source is classified as a galaxy by SDSS, then its PSF counts in $griz$ bands are obtained and compared with the corresponding flux threshold PSF counts we calculated. If the galaxy's PSF counts in the $g$, $r$, and $i$ bands are lower than the corresponding thresholds, then the galaxy passes through the mask. Otherwise, the galaxy is masked in each of the four color bands.

There is the chance that in the reference QSO imaging, faint galaxies documented by the object catalogs will pass under the flux-limited mask thresholds. Zibetti et al. (2007) used a $\sigma$ clipping procedure to exclude these faint sources. However, in a large sample of images, the random distribution of these faint background galaxies in the reference QSOs is also present in the absorber QSOs. Therefore, when the reference QSO stacking is subtracted from the absorber QSO stacking, the excess residue contributed by these faint background galaxies is negated. As demonstrated best by the $r$ and $i$ band QSO stackings, there is random residual at large impact parameters in the reference frame as well as the absorber frame, but the net frame indicates significant residue at lower impact parameters.

After masking, the mean background value measured between 400 and 500 kpc from the QSO's centroid is subtracted from the image. This same background subtraction procedure is used for every PSF star selected, as described in Section 3.4. In order to make sure that the background calculation is not contaminated by significantly brighter sources that pass under the masking algorithm, we employ a $\sigma$ clipping procedure with a threshold of $\pm3 \sigma$ and 5 iterations to exclude bright pixels.

\begin{figure*}
	\includegraphics[width=\columnwidth]{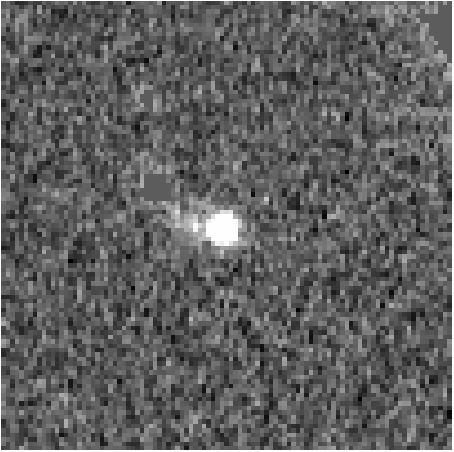}
    \includegraphics[width=\columnwidth]{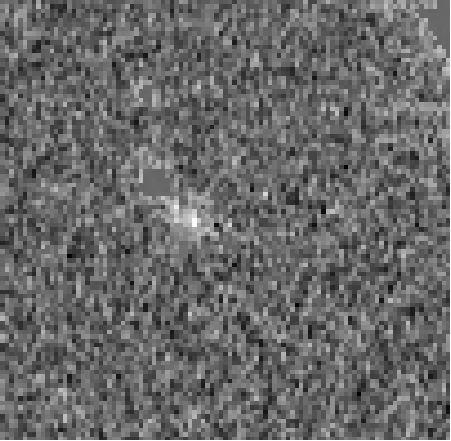}
  \caption{Top left the $r$ band cutout of QSO $J153858.91+135850.6$. Right is the PCA generated PSF subtracted image of the same quasar. The two images are shown through the Z-scale setting in SAOImage DS9. (Version 7.5) The blotches of uniform gray are sources that were effectively masked by suppressing the bright pixels to the $\sigma$ clipped mean value. There is a clear bright patch near the center of the PSF subtracted frame, which is likely the absorber host galaxy. Also note that the two frames appear almost geometrically identical: both bright and dark spots correspond almost identically in the two frames, demonstrating the low amount of noise added during PCA-generated PSF subtraction. }
  \label{fig:Figure 3.}
\end{figure*}


\subsection{QSO PSF Subtraction}

Because a quasar's light can spread to a few arcseconds and contaminate the absorption host galaxies surface brightness, the quasar's profile must be subtracted. Originally, like Zibetti et al. (2007), we utilized single point source PSF subtraction algorithm. However, upon closer analysis of individual PSF subtracted frames, single source PSF subtraction is ineffective, so we implemented a machine learning algorithm, Principal Component Analysis (PCA), to fit the QSO profiles.

At first, we chose one bright, unsaturated star as the PSF by comparing Full Width Half Maximum (FWHM), brightness, and color differences from the respective properties of the quasar. The PSF stars are normalized to the quasar using the PSF magnitudes given in the QSO catalog, then subtracted from the absorption and reference quasars. Theoretically, subtracting the PSF source from the absorption quasars should yield residues, which are the absorption host galaxies. Subtracting the PSF source from the reference quasars should yield mostly noise, as there are no absorbers. However, a substantial portion of the reference quasars left bright residue after PSF subtraction. Previous studies indicate that this kind of ``additive excess" (Zibetti et al. 2007) is caused by randomly distributed foreground galaxies in the line of sight towards the QSOs as well as by host galaxies associated with the QSOs themselves. Regardless, our single source PSF subtraction algorithm was insufficient for subtraction especially since the availability of closely related stars in each frame was low.

From a statistical standpoint, single source PSF subtraction yields greater noise. Since the single source PSF and surrounding background should closely match the quasar, the final $\sigma$ is approximately 1.4 ($\sqrt{}2$) times the original $\sigma$. The SNR of individual images is already exceedingly low, so increasing the noise by over 40 per cent is inhibitive. This is even more important with low sample sizes as with the 2DAs.
	
For those reasons, we employed Principal Component Analysis (PCA) as our means of PSF subtraction. The idea of the PCA is to perform dimension reduction on the imaging data of many PSF stars to simplify the large data sets. This method is widely used in deriving quasar continuum in the quasar spectra analysis (Francis et al. 1992; Yip et al. 2004; Suzuki et al. 2005). Instead of 1D spectra, We used PCA to analyze 2D images. For each quasar target, we select all the field stars that satisfy certain criteria to build the QSO PSF mask. Our criteria are that the PSF star must have a second moment within 10 per cent of the QSO's second moment, the PSF star can not have interpolated or oversaturated pixels, cosmic rays, its apparent magnitude must be < 18, etc. Most QSO's were matched with more than 20 PSF stars. For each field star selected, a $101 \times 101$ pixels$^2$ square cutout centered at and interpolated to the PSF star's centroid is reshaped to a single row of 10201 pixels. Each source chosen is normalized by PSF counts (given by SDSS) to the quasar's PSF counts, then put into a large array, with each row holding one source. Normalization of each source is done before the actual fitting because PCA is very sensitive to the relative fluctuation of the original variables -in this case, stars. The mean vector of the array for every pixel column is calculated and subtracted from the large array. Here we used singular value decomposition (SVD) solver to derive the ``eigen-images'' from the covariance matrix of field stars. We only used the central $13 \times 13$ eigenvectors to fit the corresponding target. The coefficients are calculated by projecting the ``eigen-images'' onto the target image. The QSO cutout and the generated PSF image, both $101 \times 101$ pixels$^2$, are then subtracted to generate the final PSF-subtracted image.

\begin{figure*}
	\includegraphics[width=\columnwidth]{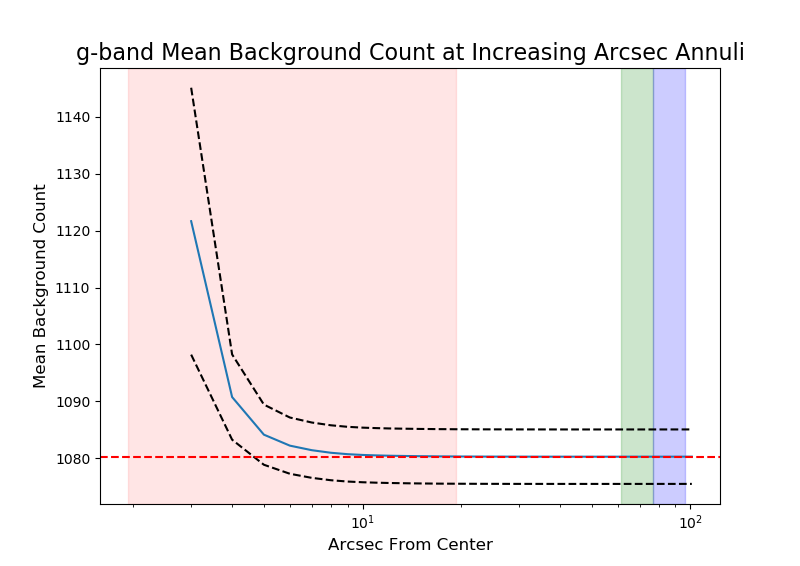}
    \includegraphics[width=\columnwidth]{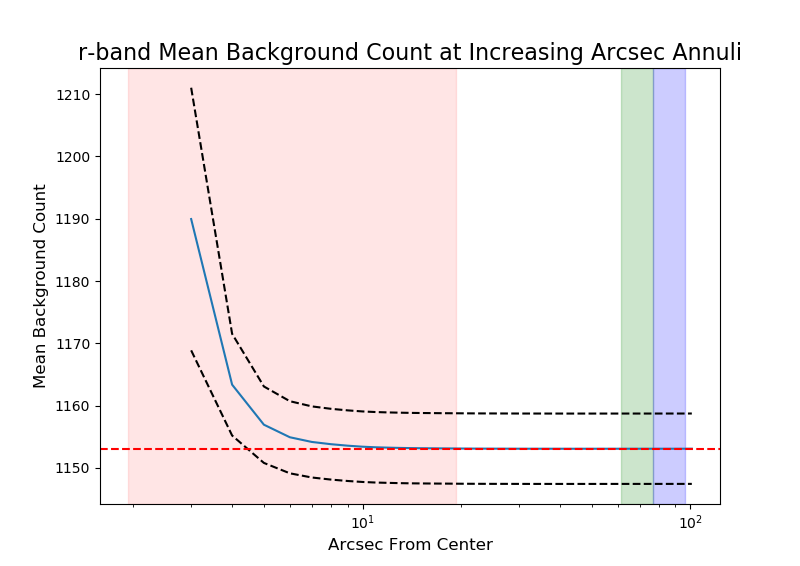}
    \includegraphics[width=\columnwidth]{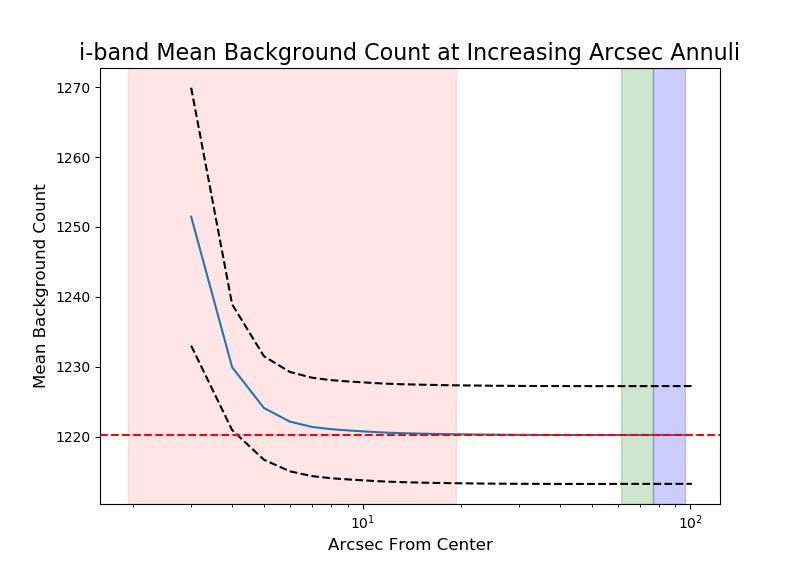}
    \includegraphics[width=\columnwidth]{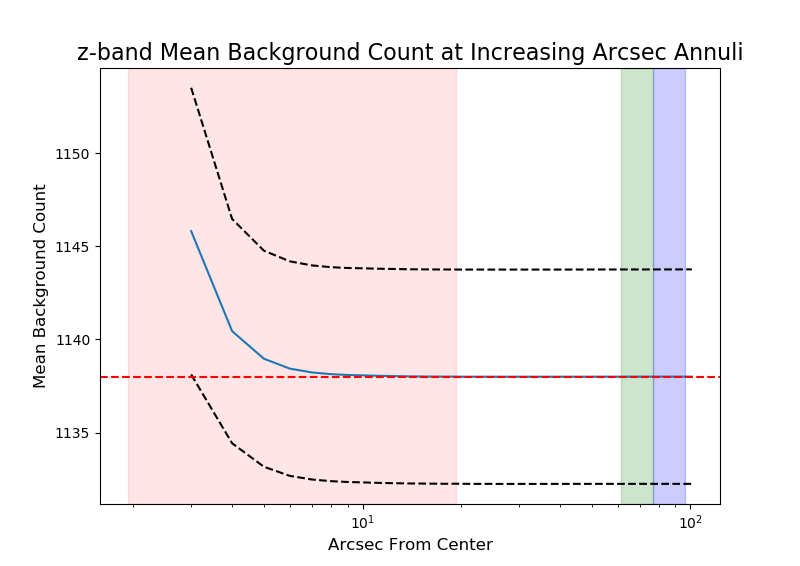}
  \caption{The average $\sigma$ clipped backgrounds of all the PSF stars used for creating the reference QSO PSF mask are plotted above (solid blue curve). A separate measurement is taken at each increasing arcsec annuli, and the mean is calculated for all PSF stars in the color band. For illustration, the red dashed line representing the average background at $400 - 500$ kpc is also plotted, again calculated as the mean of all PSF stars selected in the color band. The black curves represent the $\pm1$ $\sigma$ boundaries. The green bar represents the range that 400 kpc resides in given the redshift range of $z_{abs} = 0.37 - 0.55$, while the blue bar represents the range that 500 kpc resides in in the same redshift range. The red bar represents the range used to calculate the SED ($10 - 100$ kpc). }
  \label{fig:Figure 4.}
\end{figure*}

\begin{figure*}
	\includegraphics[width=\columnwidth]{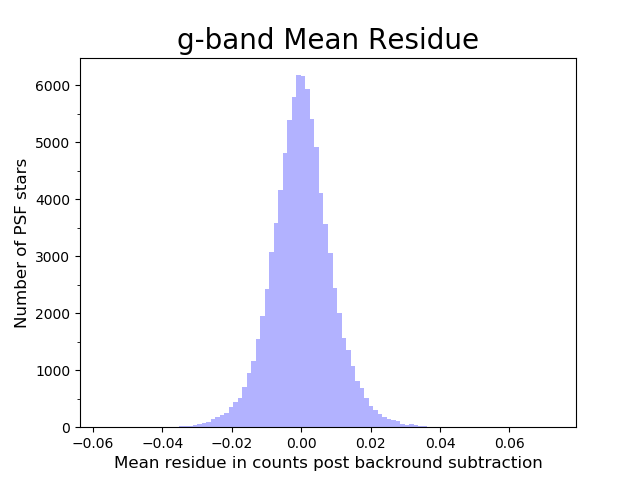}
    \includegraphics[width=\columnwidth]{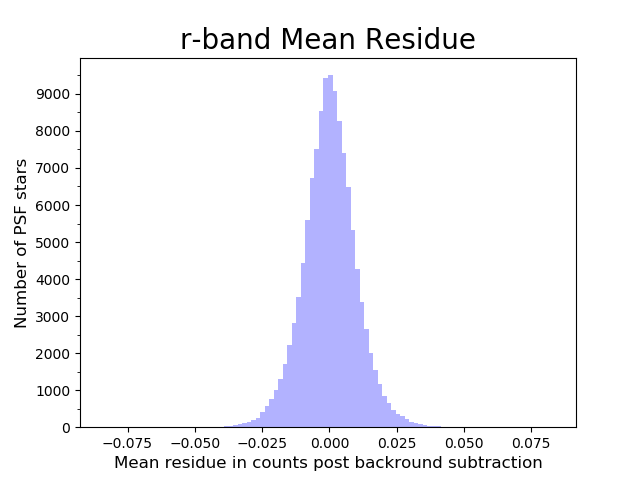}
    \includegraphics[width=\columnwidth]{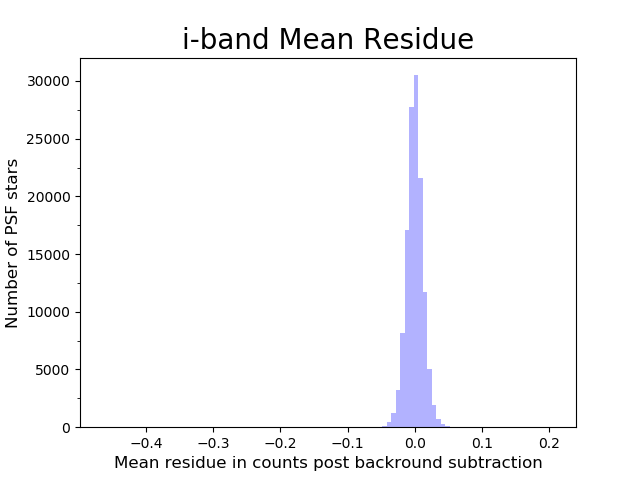}
    \includegraphics[width=\columnwidth]{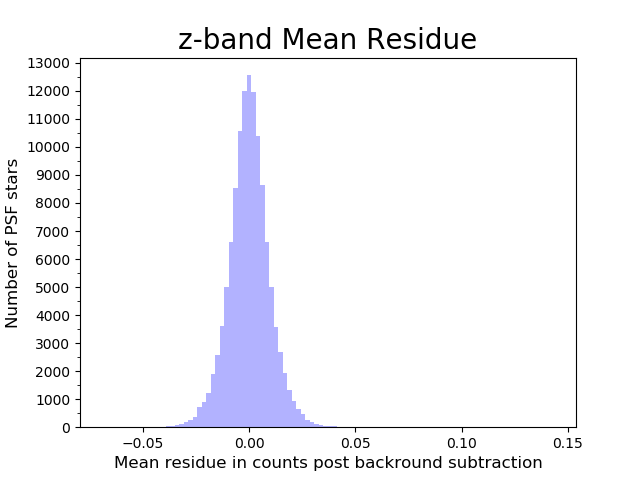}
  \caption{The mean residue of all reference QSO PSF stars at $400 - 500$ kpc after the $\sigma$ clipped background is subtracted. As shown in all four histograms corresponding to the four color bands $griz$, the mean residue left over is roughly Gaussian with a mean of 0. }
  \label{fig:Figure 5.}
\end{figure*}

\begin{table}
	\caption{Background Levels}
	\label{tab:example}
    $\sigma$ clipped mean background levels measured between $400 - 500$ kpc in counts \\
	\begin{tabular}{lccr}	
  		\hline
  		$g$ & $1080.30 \pm 4.78$\\
        $r$ & $1153.07 \pm 5.65$\\
        $i$ & $1220.24 \pm 7.00$\\
        $z$ & $1138.00 \pm 5.76$\\
  		\hline
 	\end{tabular}
\end{table}

\begin{figure*}
    \includegraphics[width=\textwidth]{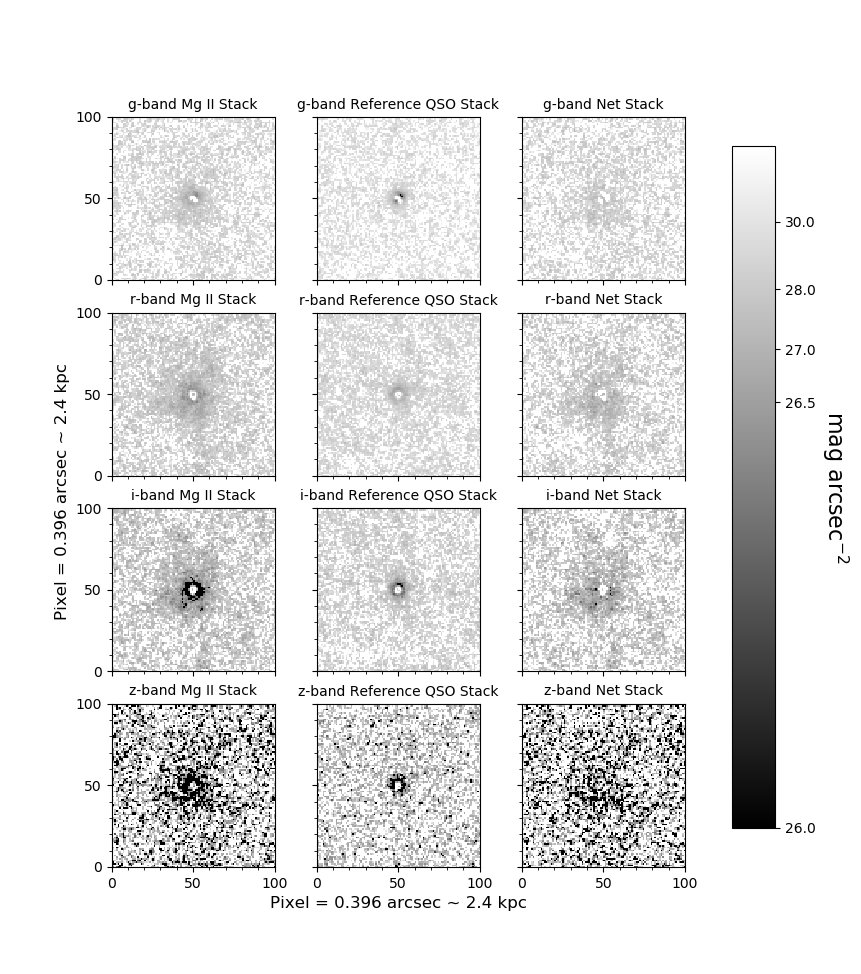}
  \caption{Above are the graphs of the four bands of PSF-subtracted images. The top row is the stacked Mg II frames, the middle row is the stacked reference QSO frames, and the bottom row is the net frame resulting from subtracting the reference stacked image from the Mg II stacked image. All pixel intensities are converted to a common mag arcsec$^{-2}$ scale. }
  \label{fig:Figure 6.}
\end{figure*}

\begin{figure*}
    \includegraphics[width=\textwidth]{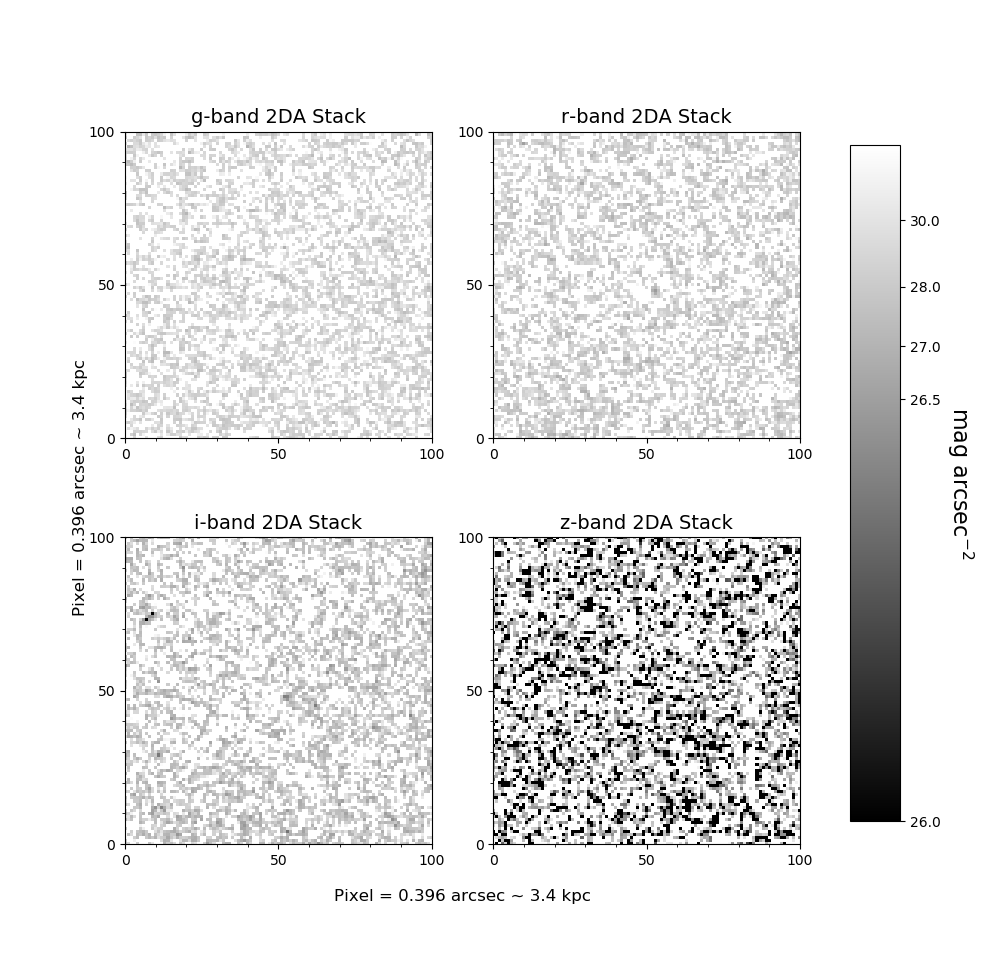}
  \caption{The stacked frames of the 2DA systems in $griz$ bands. There is little residue prevalent in each bandwidth, indicating that the 2DA host galaxies are likely at extremely small impact parameters such that they are completely subtracted off during PSF subtraction. In all four bands, the Mg II absorbers clearly contain much more absorbed light than the 2DAs. }
  \label{fig:Figure 7.}
\end{figure*}

We used Incremental PCA (IPCA), a more memory-efficient solution than regular PCA. IPCA works similarly to PCA with the notable exception that it does not store the entire dataset in memory, making IPCA advantageous over regular PCA for large datasets, as in our case. 
	
One drawback of PCA is overfitting. If too many noise components are included in the fitting calculation, then the fitting is too accurate. In addition, host galaxies close to the absorption quasars may be fitted and subtracted, leaving no visible residue. The solution was to only perform PCA on the central portion of the quasar, roughly $13 \times 13$ pixels$^2$ square portion. The coefficients used in building the PCA mask are computed using only the central $13 \times 13$ pixel$^2$ eigenvalues. The resulting PCA PSF mask is the same size as the QSO frame ($101 \times 101$ pixel$^2$). This greatly reduced overfitting while ensuring that a good profile fitting was produced. 
	
Again, from a statistical standpoint, using multiple sources yields a much smaller addition to the noise. The combined $\sigma$ of multiple sources decreases by a factor of $\frac{1}{\sqrt{N}}$ which means the more sources used, the smaller the $\sigma$, and therefore the smaller the noise added to the overall image. As shown in Figure 3, the QSO's profile is accurately subtracted from the frame, and the resulting images are geometrically similar because the noise level is significantly decreased by $5 - 6$ times.

In order to test the effectiveness of PCA PSF subtraction, we first test model star profiles. We select a bright star > 18 mag that is similar to the Mg II absorber QSO in light distribution (second moment), contains no interpolated pixels, is not over saturated, etc. We then apply the PCA PSF subtraction procedure detailed above to build a PSF mask for the chosen star, taking care to avoid including the star itself in the PSF star sample used to build the PCA PSF. The PSF subtracted frames are stacked (see Section 3.4 for more information) and the resulting mean stack frame is plotted in gray scale. Some images are excluded because there was a bright source nearby or insufficient PSF stars were found to build the PCA PSF mask. Apart from random background photon noise, there is no significant residue left over, thus demonstrating the reliability of PCA PSF subtraction.

\subsection{Photometric Calibration and Image Stacking}

Due to the nature of PCA fitting and the size constraints of the images, it is extremely difficult to perform a second round background subtraction using the method detailed above (taking the $\sigma$ clipped mean background between 400 and 500 kpc from the QSO centroid. In order to ensure that the initial background subtraction is accurate and the resulting ``pedestal" left over is minimal, we measure the $\sigma$ clipped mean background of each reference QSO PSF star (selected as described in Section 3.3) before PCA PSF subtraction and background subtraction but after masking at increasing arcsec annuli. No photometric calibration for individual images (as described in the following paragraphs) is performed in this procedure. We then take the mean count of all the background counts at each arcsec annuli for all four color bands. In the individual frames, the $\sigma$ clipped mean background varies $< 1 \sigma$ from $\sim 20$ arcsec and above, with slight random fluctuations. The background calculated between 400 and 500 kpc is therefore very representative of the true background, as shown in Figure 4. The $\sigma$ clipped mean background and $1 \sigma$ counts are given in Table 1. 

For a more thorough analysis of the level of residue left over in individual frames, we calculate the mean count between 400 and 500 kpc after the $\sigma$ clipped background level is subtracted. The resulting histograms are plotted in Figure 5. All four color bands show a Gaussian distribution of the residue with a center of 0. Thus, a second round background subtraction is highly unnecessary. 

Because the photometric calibration of each image is different, i.e. different counts correspond to different magnitudes, the images must be intensity rescaled to a uniform calibration. Each quasar is then de-reddened using the Schlegel, Finkbeiner \& Davis 1998 (SFD) extinction maps, as galactic extinction must be corrected for to ensure photometric accuracy; the resulting images are slightly brighter. The formula used is:

\begin{equation}
  I_{cal}=I_{raw}\frac{f_{20,ref}}{f_{20}}10^{0.4A_\lambda}.
	\label{eq:1}
\end{equation}

Where $I_{cal}$ is the final calibrated intensity (Fukugita et al. 1996; Smith et al. 2002), $I_{raw}$ is the original intensity, $f_{20,ref}$ is the counts that corresponds to 20 mag in the final stacked frame, $f_{20}$ is the pixel counts that corresponds to 20 mag in the observed fpC frame, and $A_\lambda$ is the galactic extinction coefficient in the color band of the quasar, given in the quasar catalogs.

\begin{figure*}
	\includegraphics[width=\columnwidth]{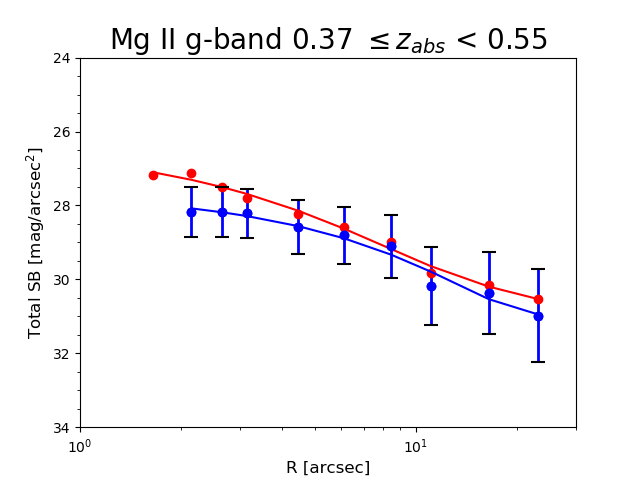}
    \includegraphics[width=\columnwidth]{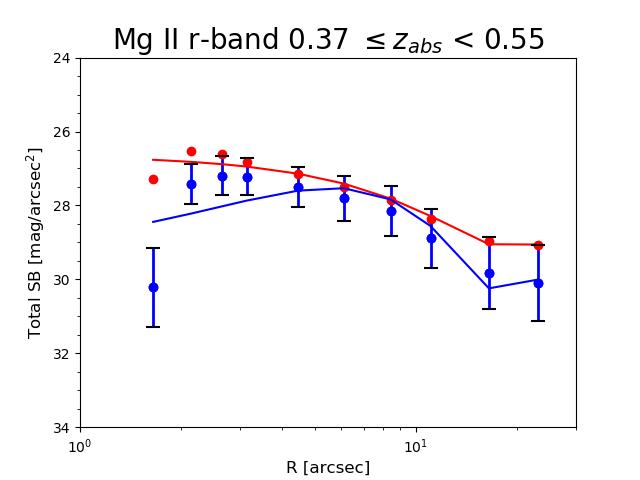}
    \includegraphics[width=\columnwidth]{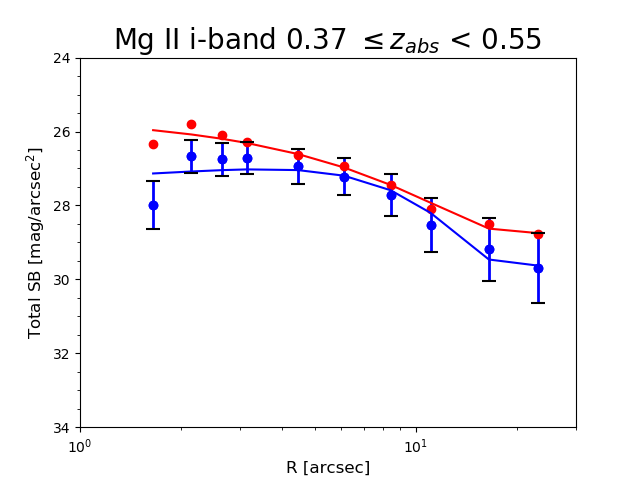}
    \includegraphics[width=\columnwidth]{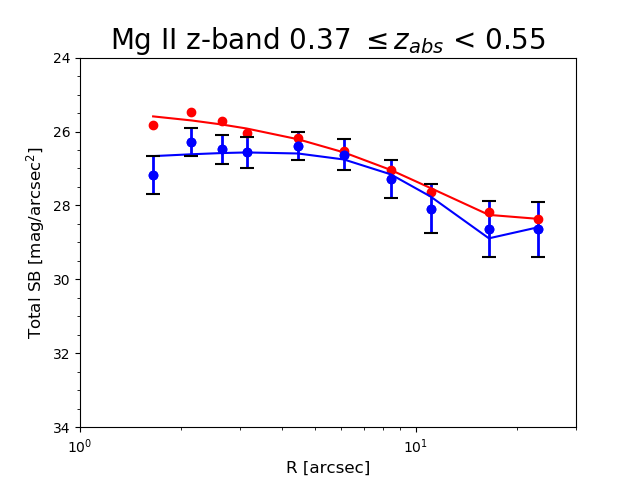}
  \caption{Above are the SB profiles of the four bands: $g$, $r$, $i$, and $z$, where most of the flux originating from the absorbers lie in. The points represent the larger radius of each circular annuli, with the smaller radii being the next lower respective point. Error bars are calculated using SDSS photon count error calculations and intrinsic image noise values. The data points represent the average net residual at each impact parameter.}
  \label{fig:Figure 8.}
\end{figure*}

\begin{figure*}
	\includegraphics[width=\columnwidth]{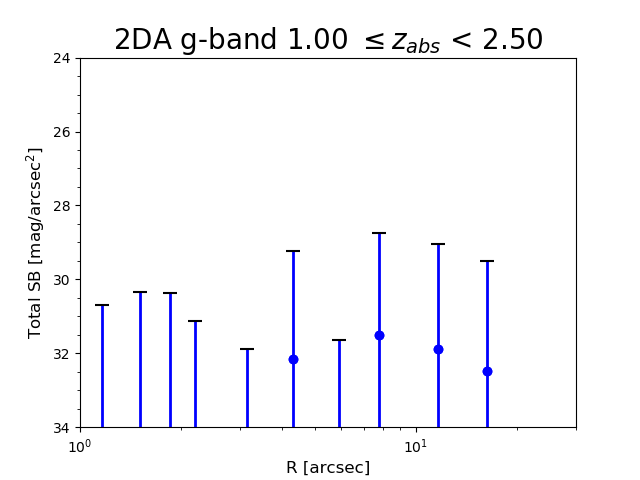}
    \includegraphics[width=\columnwidth]{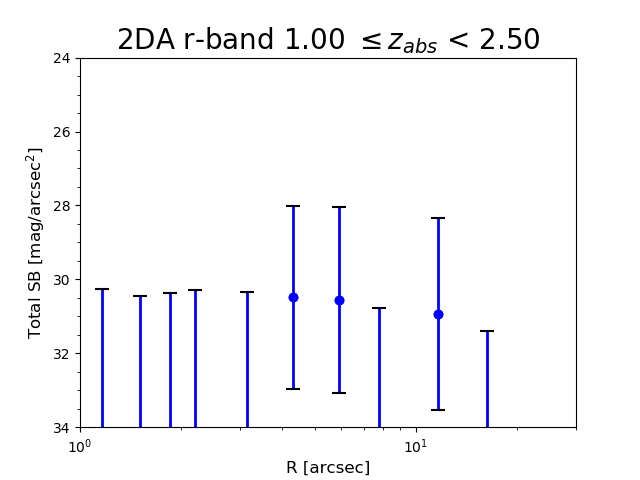}
    \includegraphics[width=\columnwidth]{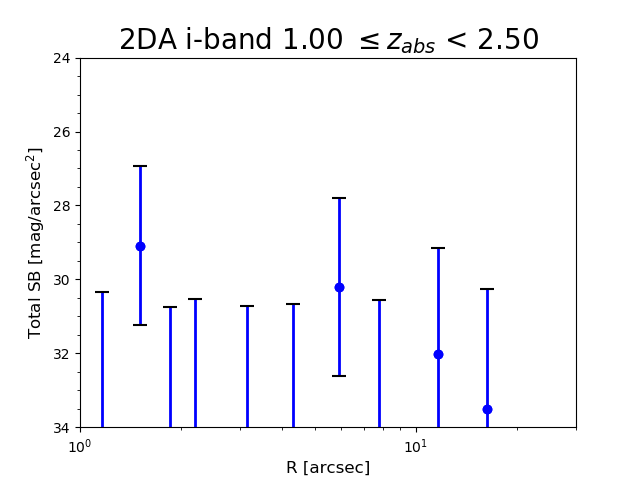}
    \includegraphics[width=\columnwidth]{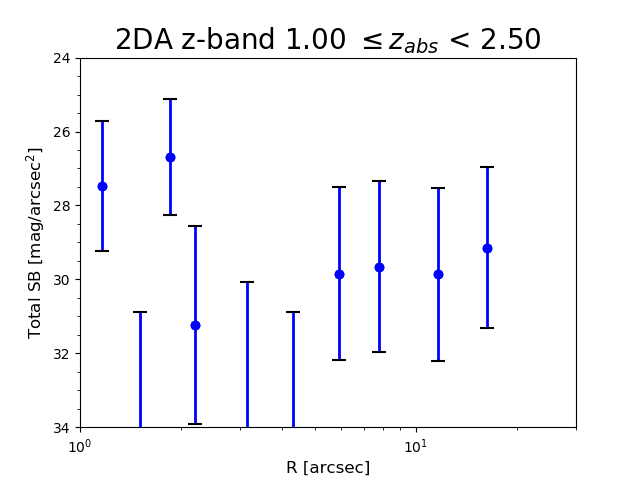}
  \caption{Same arrangement as Figure 8, except for the $g$, $r$, $i$, and $z$ bands for 2DAs. Note that no reference QSOs were subtracted from the stacked 2DA frame. Very little residue is detected in all four bands.}
  \label{fig:Figure 9.}
\end{figure*}

The images are then combined by taking the mean value of each pixel value separately. Since the host galaxies are clustered around the quasars, once all the PSF subtracted images are stacked, a significant halo is observed. This halo is the combined residue of all the absorption host galaxies, which were mostly undetectable in individual frames.

Unlike in Zibetti et al.'s (2007) study, we do not physically rescale the images to a common kpc pixel$^{-1}$ scale. This is because the location dependent sky background is not accurately subtracted from the entire image; only a $\sigma$ clipped mean value is subtracted. Although the average sky background has been greatly suppressed as demonstrated in Figure 4 and 5, localized sky level fluctuation still exist. Thus, physical rescaling will result in flux deviations. The purpose of such rescaling was to ensure different host galaxies corresponding to different redshifts would have the same kpc pixel$^{-1}$ scale. Without rescaling, the final stacked images would become slightly ``blurred" as residues from different physical impact parameters are combined. However, due to the low range of redshifts ($z_{abs} = 0.37 - 0.55$), this was deemed acceptable; ultimately, when calculating total flux and SED, the ``blurring" effect is minimized since the cumulative flux of the final stacked image is calculated. For calculating the SB profiles and fitting the SED, we use the mean $z_{abs} \sim 0.48$.

All the processes performed are created from a package developed by us, implementing standard libraries such as Numpy, Scipy, Astropy, Photutils, sklearn, etc. The Atlas.fits files that contain all the ``poststamp'' deblended images of each individual object is processed through the sdsspy Python package which we modified and implemented. The flux-limited masks thresholds are calculated by implementing the EzGal Python package developed by Mancone et al. (2012), which utilizes the $griz$ band fluxes from SDSS and the stellar population SEDs calculated from Charlot \& Bruzual (2007). A great majority of the processing code is written in Python 3. The data download from the SDSS Data Archive Server is written in Batch programming.

\begin{figure*}
    \includegraphics[width=\textwidth]{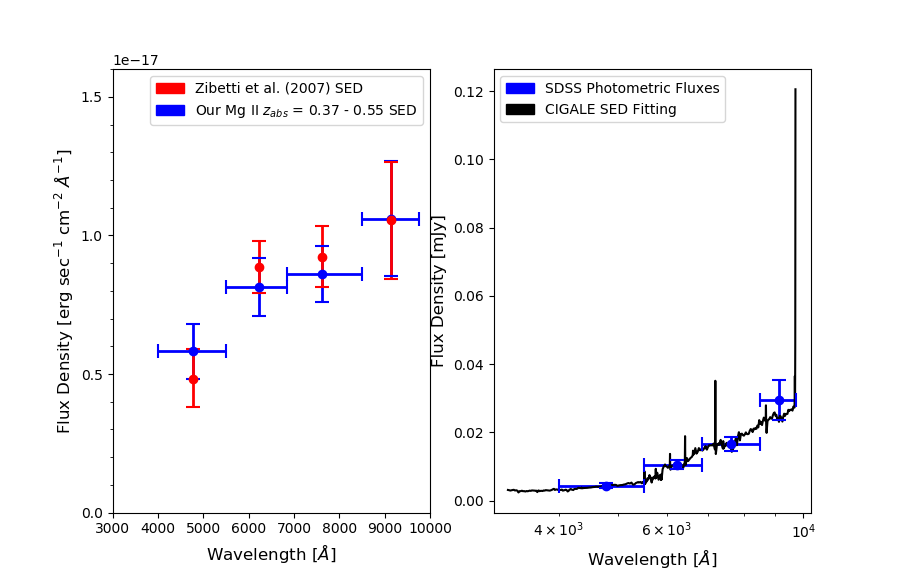}
  \caption{The SED plots of the Mg II absorber Host Galaxies (Left) with error bars, calculated from intrinsic photon noise and sky background noise. The magnitudes shown in the left plot are calculated by converting the integrated flux between 10 kpc and 100 kpc to magnitudes using SDSS zero point flux densities. Zibetti et al.'s (2007) SED for $0.37 \leq z_{abs} < 0.55$ Mg II absorbers is also plotted. Note the remarkable consistency between our derived SED and Zibetti et al.'s (2007). The right plot is the best SED fitting generated using the CIGALE software described in Burgarella et al. (2005). The resulting $\chi^2$ fitting is 0.67. }
  \label{fig:Figure 10.}
\end{figure*}

\section{Photometric Analysis}

In this section we analyze the statistical properties of the stacked frames. In Section 4.1, we present the SB profile of the Mg II absorber host galaxies. In Section 4.2, we plot the Spectral Energy Distribution (SED) of the Mg II absorber host galaxies. In Section 4.3, we simulate the upper limit threshold of the average impact parameter of the 2DA host galaxies.

\subsection{Surface Brightness (SB)}

Despite our best fitting, there is still some residue left over from the PSF subtraction in the reference quasar frames as shown in Figure 6. This residue is demonstrated in the appendix section by Zibetti et al. (2007) as a systematic effect of the redshift and magnitude of the QSO itself. As such, this is taken as the ``additive excess" of the quasars. The stacked reference quasar residue is subtracted from the stacked absorber quasar reside to yield a net profile. We calculate the SB at annuli with geometrically increasing radii in order to preserve a near-constant SNR. Figure 7 shows the residue left from the 2DA stacked frames. Figure 8 shows the radial distribution of SB profiles at $g$, $r$, $i$, and $z$ bands for Mg II absorber host galaxies, and Figure 9 shows the SB profiles for 2DA host galaxies. 

It is worth noting that increasing redshifts lead to increasing physical scales of kpc pixel$^{-1}$. Because we did not physically rescale the data to a constant kpc pixel$^{-1}$ scale, our image stacking method results in slight blurring of absorber host galaxies at different redshifts. In addition, due to the normalization of the PSF to the QSO, the center is zeroed out to a high degree, indicative of the good quality of the PSF fitting. However, this also results in any low impact parameter host galaxy's light being subtracted off. For Mg II absorbers, this is not a problem since there are very few, if any, host galaxies associated with Mg II absorbers at an impact parameter of < 10 kpc, or roughly $4 - 5$ pixels (Steidel et al. 1994; Churchill et al. 2005a). Regardless, there is significant residue left over in the net profile.

The SB profiles for the 2DA host galaxies in the $griz$ bands are also given in Figure 9. From the plots, it is clear that there is no significant residue left over in all four bands. As shown in Figure 7, there appears to be no significant residue. Therefore, it is impossible to fit power laws or derive an SED.

\subsection{Spectral Energy Distribution (SED)}

We then built an SED of the Mg II absorber host galaxies. The total flux of the absorbers from $10 - 100$ kpc is integrated and converted to flux density units using the zero-point magnitudes given by SDSS. The magnitudes obtained are consistent with Zibetti et al.'s (2007) results as illustrated in Figure 10, further confirming the success of the PCA fitting algorithms.

We then proceed to use the CIGALE Bayesian SED fitting library developed by Burgarella et al. (2005). This library fits the SED in the observed frame; consequently no K-correction is necessary. Compared to the galaxy template fitting used in Zibetti et al. (2007) that can only indicate the general type of galaxy, CIGALE can derive the physical parameters of Mg II absorber host galaxies. In order to fit the SED of Mg II photometry data we derived, we had to decide many of the components used to fit the data. We used the delayed $\tau$ model for the star formation history (Lee et al. 2009). We also used the models provided by Bruzual \& Charlot (2003) to model the stellar population evolution. York et al. (2006) derived the mean extinction curve by geometrically combining Mg II absorber spectra and determined that there was an absence of the 2175 \AA\ absorption bump. Therefore, we used the SMC extinction curve to model the dust attenuation of Mg II host galaxies. According to the wavelength coverage of SDSS, the imaging data does not extend to the infrared region. Thus, no dust emission component is included in the Mg II SED modeling. Here we used the $z_{abs} = 0.48$ mean redshift of our Mg II absorber sample as the redshift of the Mg II SED. Since the $z$-band data is very close to the $H\alpha$ emission in the Mg II rest frame, we add the nebular emission into the SED modeling. Because the average redshift of the Mg II absorbers is 0.48, the oldest age of stars in absorber galaxies is around 8.750 Gyr; thus, the highest age input for the oldest star parameter is 8.750 Gyr. We carefully tuned the parameter ranges for each of the components, including metallicity and dust attenuation power law, and the best fitting we obtained produced a $\chi^2$ of 0.67. 

Because of the uncertainty of the range of physical parameters that would result in a best fit, we inputted a wide range for the parameters, including maximum star age, best slope delta of the power law continuum, V-band attenuation of the young star population, etc. The Bayesian parameters and errors given are averaged over the posterior distributions of the output parameters, while the best fit parameters are reported from the best $\chi^2$ fitting. We report both sets of parameters in Table 2. Table 3 provides the apparent magnitudes derived from the integral flux between $10 - 100$ kpc. 

Some of the best fit parameters diverge from the Bayesian parameter i.e. stellar mass. This is most likely due to the low wavelength range of our data which does not extend into the UV range. 


The best fit (lowest $\chi^2$) star formation rate we derived here is 2.2 $M_{\odot}$ $yr^{-1}$. The star formation rate of Mg II absorbers are usually derived from the nebular emission lines in the spectra data. Noterdaeme et al. (2010) identified 46 Mg II spectra in the redshift range $< 0.8$ with [OIII] emission lines from SDSS DR7. The derived lower limits of SFR is in the range of $0.2 - 20$ $M_{\odot}$ $yr^{-1}$, which is consistent with our result. Recently, Joshi et al. (2017) derived the star formation rate (SFR) of Mg II in the redshift range from 0.35 to 1.1 by measuring [OII] luminosities from 198 spectra in SDSS DR7 and DR12. The SFR is in the range $1 - 20$ $M_{\odot}$ $yr^{-1}$, which is also consistent with our derived best fit SFR of 2.2 $M_{\odot}$ $yr^{-1}$.

No SED is produced for the 2DA stacked frame because the amount of residue left is insignificant; this implies that the average impact parameter of 2DA host galaxies is either much smaller than that of Mg II absorber host galaxies or the signal is still much too low for detection in the stacked frame. We will explore this in the next section.


\begin{table}
	\caption{The best fitting produced by CIGALE}
	\label{tab:example}
    Average Mg II Absorber Host Galaxy Parameters $z_{abs} = 0.37 - 0.55$\\
	\begin{tabular}{lccr}	
  		\hline
  		Best $\chi^2$ Star Formation Rate (SFR) & $2.2$ $M_\odot$ $yr^{-1}$\\
  		Best $\chi^2$ Stellar Mass & $4.0 \times 10^{10}$ $M_\odot$\\
       	Best $\chi^2$ Av & $0.14$\\
        Bayesian SFR & $2.3 \pm 1.0$ $M_\odot$ $yr^{-1}$\\
        Bayesian Stellar Mass & $(2.4 \pm 1.1) \times 10^{10}$ $M_\odot$\\
        Bayesian Av & $0.33 \pm 0.23$\\
  		\hline
 	\end{tabular}
\end{table}

\begin{table}
	\caption{Integral Photometry Magnitudes}
	\label{tab:example}
    Integral flux from $10 - 100$ kpc converted to apparent magnitudes
	\begin{tabular}{lccr}	
  		\hline
  		$g$ & $22.29\SPSB{$+$0.20}{$-$0.17}$\\
        $r$ & $21.34\SPSB{$+$0.15}{$-$0.13}$\\
        $i$ & $20.84\SPSB{$+$0.14}{$-$0.12}$\\
        $z$ & $20.23\SPSB{$+$0.23}{$-$0.20}$\\
  		\hline
 	\end{tabular}
\end{table}

\subsection{2DA Impact Parameter Simulation}

The over subtraction in the 2DA data points to an inherent weakness of the stacking method, and particularly of the PSF normalization section; if the impact parameter of the absorber host galaxies is extremely low, then the stacking method would likely only yield a upper limit for impact parameter calculations and a lower limit for SED fitting. Thus, we perform simulations to test the upper limits of the average 2DA host galaxy impact parameter.

It should be noted that since 2DA systems possess large rest-EW of the 2796 \AA\ absorption line, they are strong Mg II absorber systems. Previous studies have demonstrated that the larger the EW of $W_0(\lambda2796)$, the smaller the impact parameter (Zibetti et al. 2007; Nielsen et al. 2013). Zibetti et al. (2007) showed that Mg II absorber systems with a $W_0 \geq 1.58$ have an average luminosity-weighted impact parameter of 43 kpc, while weaker systems ($0.8 \leq W_0 < 1.12$) have an average luminosity-weighted impact parameter of around 60 kpc. In addition, their redshift binning indicated that redshift does not significantly affect the impact parameter distribution. 2DA systems possess an average $W_0 = 1.57 \pm 1.16$ (Zhao et al. 2018 in prep), which is on the strong side of the Mg II absorber systems. Thus, it is expected that 2DAs have smaller impact parameters.

\begin{figure*}
	\includegraphics[width=\textwidth]{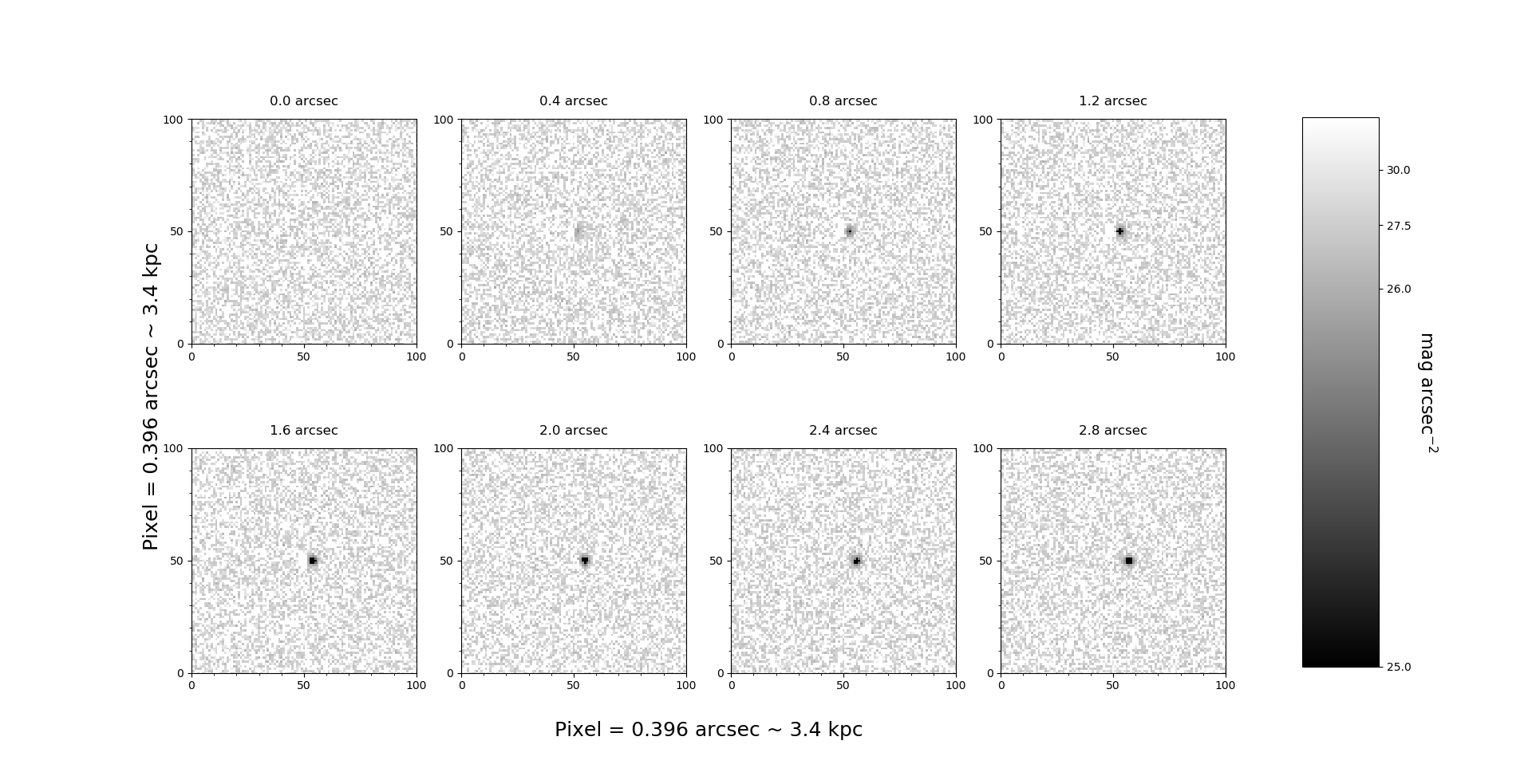}
  \caption{This plot displays the net amount of residue present at different arcsec impact parameters of the simulated absorber host galaxy. The ``seeing'' FWHM is 1.4 arcsec. In order to keep noise introductions to a minimum, only unit pixel shifts of the galaxy from the centered QSO are performed, which translates into impact parameter increments of $\sim$0.4 arcsec. The color bar has units of mag arcsec$^{-2}$ to clearly illustrate the differences in residue. }
  \label{fig:Figure 11.}
\end{figure*}

\begin{figure*}
    \includegraphics[width=\textwidth]{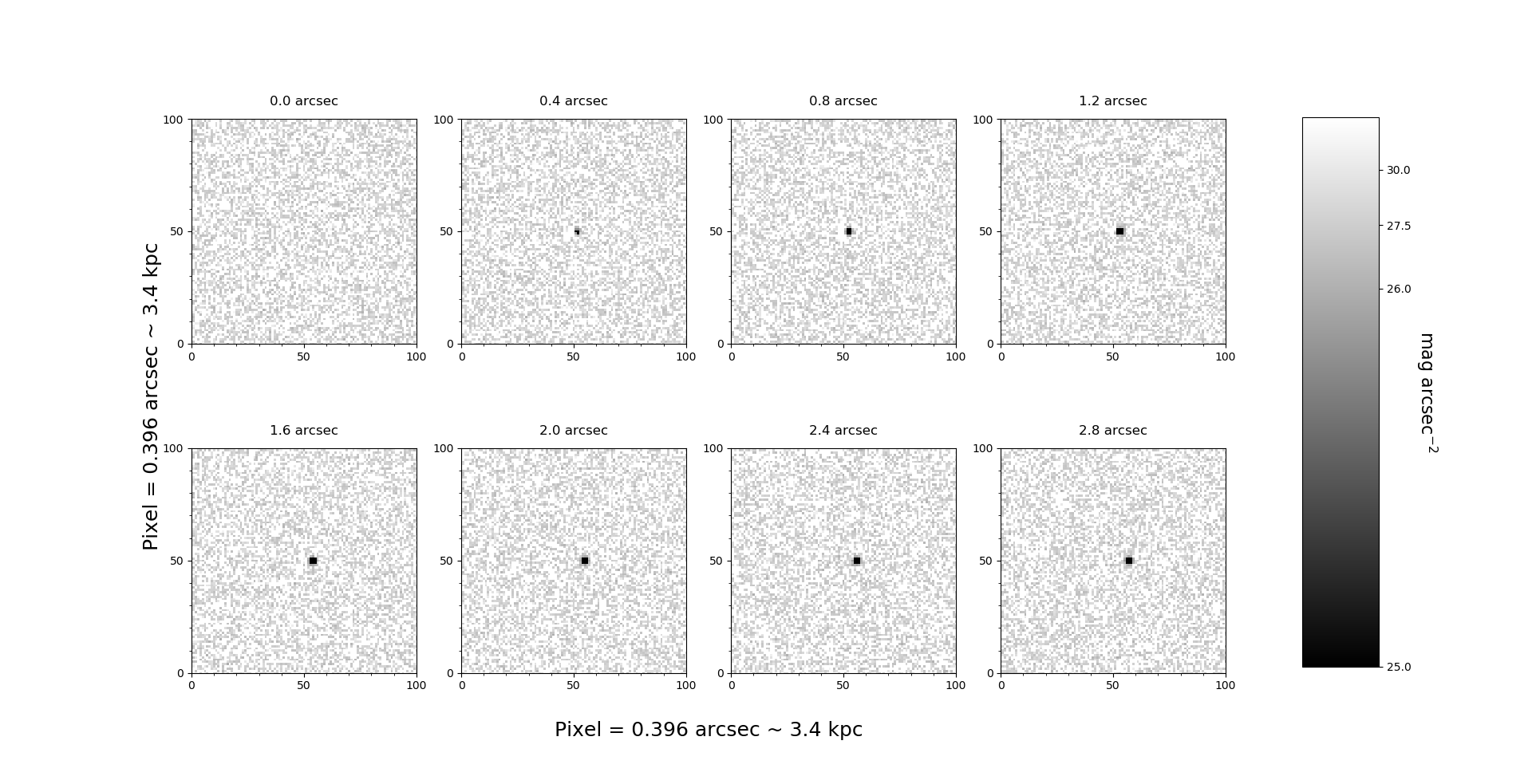}
  \caption{Same as Fig. 9 except with a seeing FWHM of 1.0 arcsec. Note that the residue becomes more pronounced at 0.8 arcsec with this particular seeing than with a seeing FWHM of 1.4 arcsec.}
  \label{fig:Figure 12.}
\end{figure*}

\begin{figure*}
    \includegraphics[width=\textwidth]{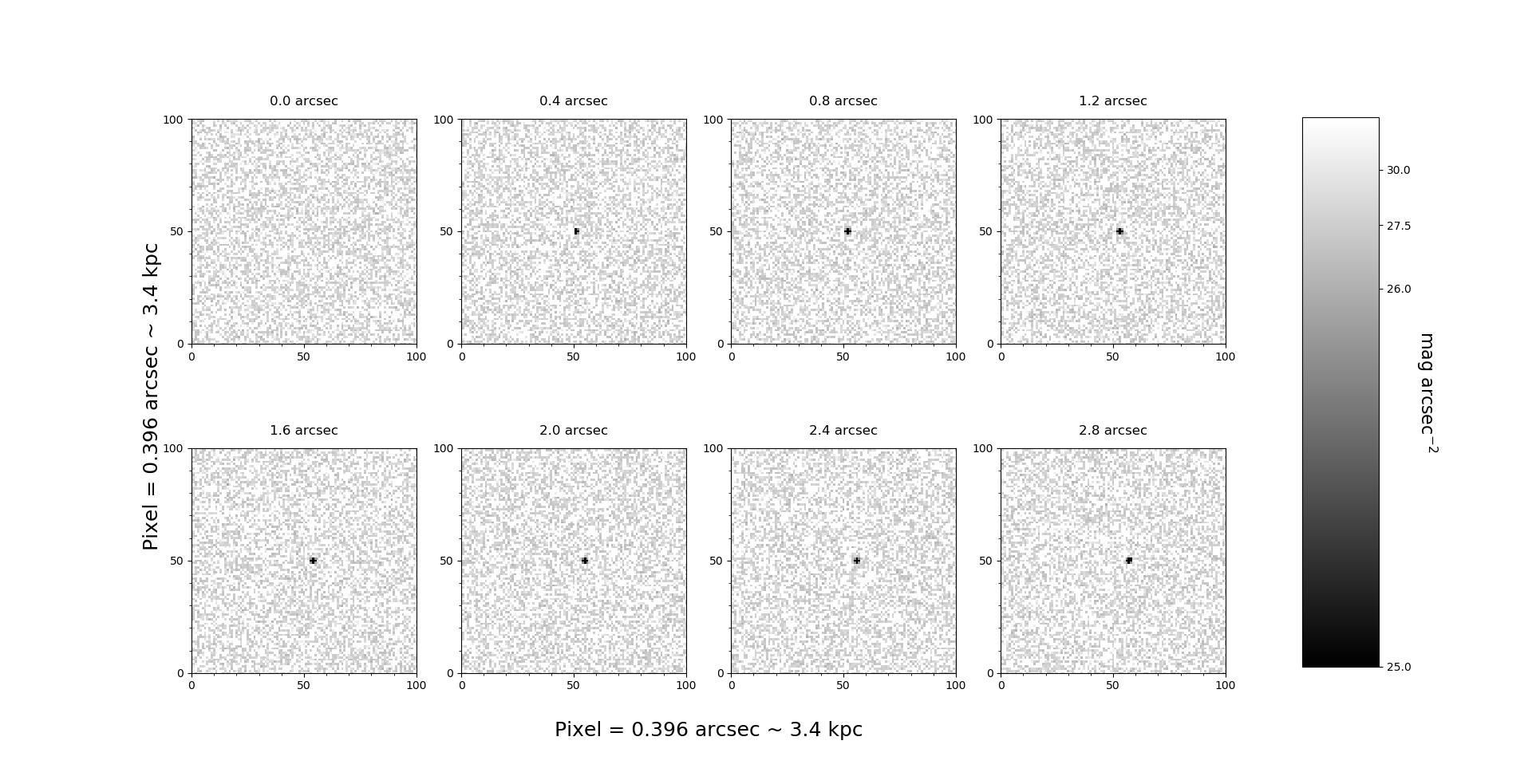}
  \caption{Same as Fig. 9 except with a seeing FWHM of 0.6 arcsec. At 0.8 arcsec, the amount of residue is much more significant. Even at 0.4 arcsec there is a significant level of detection.}
  \label{fig:Figure 13.}
\end{figure*}

\begin{figure*}
     \includegraphics[width=\textwidth]{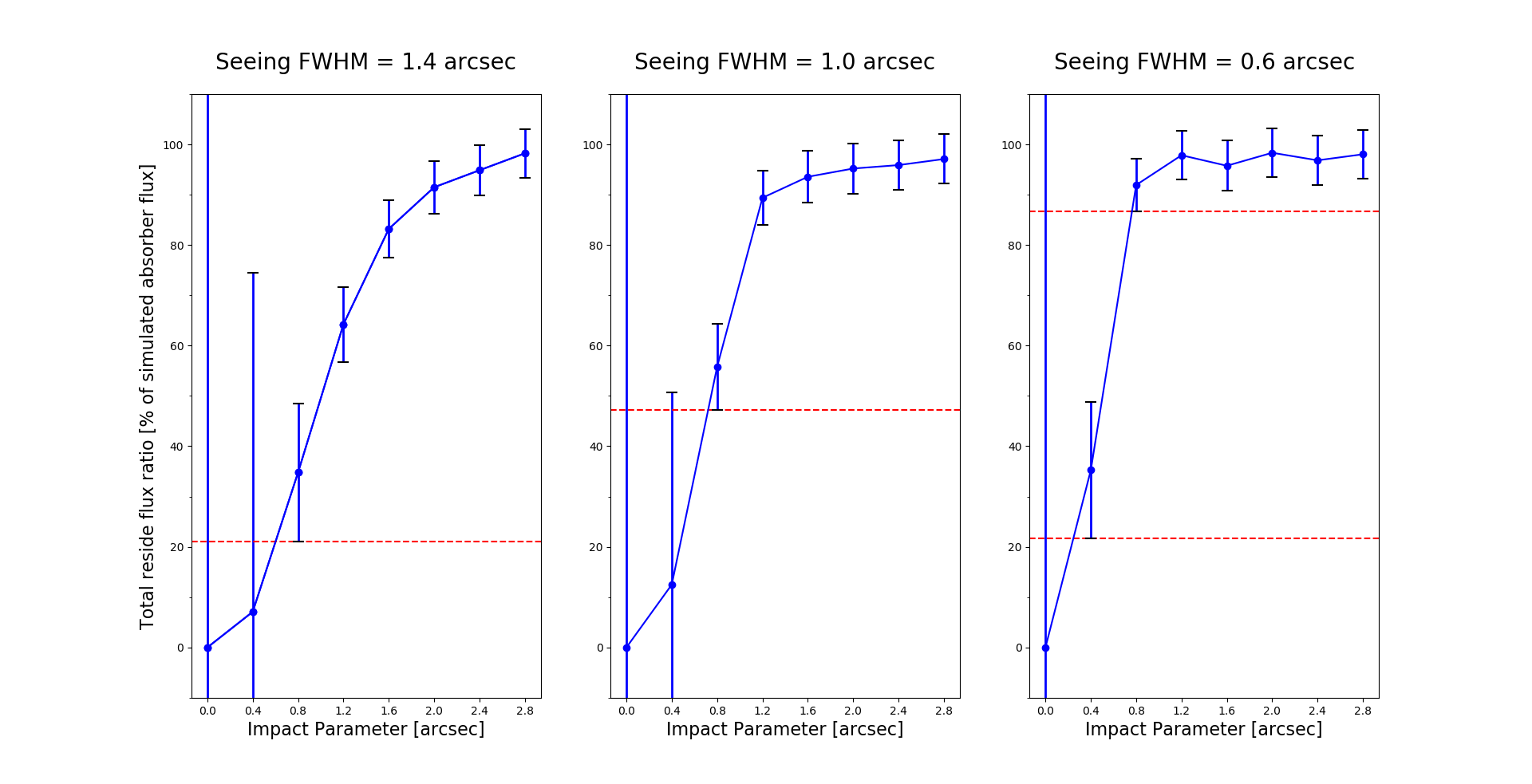}
  \caption{The above graph plots the total flux of the residue present at each impact parameter as a percentage against the simulated absorber's flux. Error bars calculated from the intrinsic photon noise and at 3 $\sigma$ are also plotted in both directions. For clarity, horizontal red lines are plotted at the lower limit of the $0.4 - 0.8$ impact parameters. }
  \label{fig:Figure 14.}
\end{figure*}

To test the upper limits of the 2DA's luminosity-weighted impact parameters, we use the entire 2DA reference quasar dataset (as selected in Section 2.3) of $\sim$1580 QSOs and their magnitudes. In order to accurately simulate the absorbing host galaxies impact parameter without these interfering non-absorbing host galaxies, we note that both ``clean'' QSOs and stars are well-approximated by PSFs, hence Gaussian profiles. In this simulation, we use an average ``seeing'' of 1.4 arcsec to produce Gaussian profile approximations of both the QSOs and the PSF stars. SDSS ``seeing'' is standard at around 1.4 arcsec, with slightly varying PSF FWHMs reported in the object catalogs, ranging from $1.2 - 1.6$ arcsec. We choose the average of 1.4 arcsec in our simulation. For the absorbers, the profiles are scaled to the magnitudes of the reference QSOs for consistency, and a second Gaussian profile with a cumulative area $= 23.7$ magnitudes in the $r$-band is overlaid at different impact parameters (0.4 arcsec, 0.8 arcsec, etc.) to simulate a 2DA host galaxy. This magnitude choice is derived from Zibetti et al.'s (2007) absolute magnitude converted to an apparent magnitude at a redshift $z_{abs} = 1.44$, and thus representative of the apparent magnitude of Mg II absorber host galaxies at that redshift. We choose the redshift $z_{abs} = 1.44$ because that is the average redshift of the 2DA absorbers. An almost uniform sky-background measured from SDSS DR12 field images is added to the resulting images, and photon noise is applied for each pixel. This photon noise is drawn from a Poisson distribution. For each simulated absorber, 10 PSF stars are created using the image-creation procedure with the exception that no host galaxy profile is added and they are normalized to the QSO with the absorber. For consistency, the seeing for the QSO, host galaxy, and the PSF stars are held constant. For illustration, we limit the number of host galaxies per absorber to one. PCA is used to build a PSF mask from the PSF stars, and the constructed PSF is finally renormalized to the QSO's total intensity as observed within 1 FWHM from the centroid.

As shown in Figures 11 and 14, at 0.4 and 0.8 arcsec impact parameters, the amount of residue present is insignificant (<3 $\sigma$). Above 0.8 arcsec, significant residue is observed (>3 $\sigma$). Thus, with SDSS field imaging data we can determine that beyond 0.8 arcsec, which corresponds to $\sim$7 kpc at an average absorber redshift of $z_{abs} = 1.44$, we are theoretically able to detect 2DA host galaxies in the stacked frame if they have the same magnitude as typical Mg II absorber host galaxies at the same redshift. However, the lack of any significant residue in all four color bands ($griz$) suggests that the 2DA host galaxies likely do not reside at such an impact parameter i.e. > 0.8 arcsec. On the other hand, the non-detection is highly unlikely to be caused by the faintness of the companion. We measured the sky background of the stacked image, which corresponds to 25.5 magnitude at 3 $\sigma$. Hence, the minimum brightness threshold for determining a source is 25.5 magnitudes, which is lower than the predicted magnitude of 23.7. Both Keck spectroscopy of 2DAs and HST imaging of the $z_{abs} = 2.12$ 2DA towards $J1211+0833$ show that host galaxies of 2DAs have stellar masses ranging from $10^{9}$ to $2 \times 10^{11}$ $M_\odot$ $yr^{-1}$ (Ma et al. 2018a), comparable to that of $z_{abs} \sim0.48$ Mg II absorber host galaxies (Table 2). In addition, Buat et al. (2012) showed that the global amount of dust attenuation increases with stellar mass. Since 2DA systems are highly dust attenuated, this means that 2DA host galaxies are likely much more massive than most of Mg II absorber host galaxies which do not show strong dust attenuations. It is highly likely that the average host galaxy magnitude is around 23.7 mag or even brighter in the $r$-band. 23.7 mag itself is well above the detection limit. Thus, we can conclude that these host galaxies on average must reside at an impact parameter lower than 7 kpc since any higher impact parameter should result in detection. This low impact parameter is also consistent with the HST deep imaging results of the 2DA system at $z_{abs} = 2.12$ (Ma et al. 2018b).

In order to test if seeing-limited image quality affects the detection limit of the impact parameter of the absorbers using ground-based telescopes, we repeat the simulation using a seeing limited image with FWHM of 1.0 and 0.6 arcsec. As shown in Figures 12 and 13, the amount of residue detected increases as the seeing FWHM decreases; more residue is detected in the 1.0 arcsec seeing than compared to 1.4 arcsec seeing at all impact parameters, more residue is detected in the 0.6 arcsec seeing than compared to 1.0 arcsec seeing at low impact parameters. The absorber host galaxy becomes fully deconvolved from the quasar at 2.0 arcsec in the 1.0 arcsec seeing simulation and at 1.6 arcsec in the 0.6 arcsec seeing simulation.

It is clear that in the 0.6 arcsec seeing simulation, between 0.4 and 0.8 arcsec, the amount of flux left over after PSF subtraction jumps from $\sim$35 per cent to $\sim$92 per cent. The better (smaller) the seeing FWHM is, the steeper the slope is. For the best seeing of 0.6 arcsec, we can detect a host galaxy of a dust absorber as close as 0.4 arcsec. This means that studying host galaxies associated with 2DAs using a stacking technique would likely require ground-based imaging with much better seeing than 1.4 arcsec with the SDSS. On the other hand, the actual sample size of 2DA systems is only a quarter of the sample size used in the simulation (see reference quasar - 2DA system ratio as noted in Section 2.3). This means the error is actually twice as large. Thus, improving the sample size to reduce measurement errors would also allow detection of host galaxies with impact parameters < 0.8 arcsec under the 1.4 arcsec seeing.

\section{Interpretation}

In this section we interpret our findings. The SED of Mg II absorbers follows that of intermediate spiral galaxies, consistent with Zibetti et al's (2007) findings (with slightly differing flux measurements). The stellar properties we derived in Table 2 indicate that the majority of stellar populations in these host galaxies are likely old. In the best fitting produced, the stellar age was 8.750 Gyr. This is quite possible since bright and massive host galaxies at this redshift range likely dominate in their contributions to the final subtracted image residuals (Figure 6). 

In addition, the average impact parameter of 2DAs is very low as there is no apparent residue left over after PCA PSF subtraction. This agrees with Ma et al.'s (2018b) study of one 2DA with redshift $z_{abs} = 2.12$ where they found the impact parameter to be 5.5 kpc (Ma et al. 2018b). Previous research has shown that within 10 kpc (i.e. $\sim$4 pixels in the redshift range $z_{abs} = 0.37 - 0.55$), there are no Mg II absorber host galaxies present (Steidel et al. 1994; Churchill et al. 2005a). However, because of the normalization of the PSF fitting to the QSO, the central portion is very nearly zeroed out. This is indicative of the high quality of PSF subtraction, but is somewhat problematic for low-impact parameter systems like 2DAs.

As demonstrated by our simulations, 2DA host galaxies possess a luminosity weighted average impact parameter of < 7 kpc. The low impact parameter suggests that 2DAs are likely associated with the disk components of high redshift galaxies. This appears to be consistent with Hubble Space Telescope imaging of high redshift main sequence star-forming galaxies (Kriek \& Conroy 2013; Ma et al. 2018b).

\section{Discussion}

In this study, we have shown that PCA is capable of reproducing quasar profiles to a high degree of accuracy. We now explore the differences in methodology. 

Zibetti et al. (2007) utilized a masking algorithm that covered up all sources deemed unlikely to be an absorbing galaxy at the given redshift. They used SExtractor as their baseline masking, supplemented by a flux-limited mask derived from a metal-poor stellar population. We used the fpObj files released by DR7 and DR12, which contain detailed categorizations of the sources found in each field image, ensuring that our masking masks all stars, extended sources (such as field galaxies, nebula), etc. 

Next, during the PCA fitting of the quasars, we chose to only take the components calculated in a $13 \times 13$ square around the centroid of the quasar. Although the fitting is highly accurate in most cases, some reference quasars leave significant amounts of residue after masking and PSF subtraction. The $13 \times 13$ was chosen to minimize overfitting based on the average extension of a non-absorbing quasar's immediate residues, which are calculated from their FWHM. 

Although the procedure was highly effective in retaining data (> 425 out of 436 passed in all bands, with failed cases being where the QSO was unable to be cut due to boundaries or no suitable sources were found for PCA PSF fitting), the amount of residue left over is low; the surrounding noise level is still very high. Two conclusions were made after comparison between the two absorber stackings:

\begin{enumerate}
\item  The high redshift of the 2DAs ($z_{abs} = 1 - 2.5$) means that the detection completion is very low. Assuming a standard cosmological model, magnitudes in that range increase by approximately $2 - 4$ from the magnitudes recorded at $z_{abs} = 0.37 - 0.55$ due to cosmological dimming. Thus, the SNR is already much weaker due to increased distance. Flux-limited masks that helped distinguish between Mg II absorber host galaxies and non-absorbing galaxies were not necessary for 2DAs. 
\item  The impact parameters for most of the 2DA host galaxies are small: < 0.8 arcsec or < 7 kpc from the quasar. Since the light of the QSO in SDSS imaging extends out to a few kpc, the 2DAs are almost completely precluded in the imaging data. 
\end{enumerate}

\section{Summary}

Using an independent image stacking procedure, we concluded that host galaxies of Mg II absorbers at $z_{abs} = 0.37 - 0.55$ are, for the most part, evolved intermediate spiral galaxies with lower SFR (2.2 $M_\odot$ $yr^{-1}$). Our SB profiles and integrated fluxes for the Mg II absorber host galaxies closely resemble Zibetti et al.'s (2007). This demonstrates that our stacking technique is reliable. No SED or SB profile could be derived from the 2DAs due to the low impact parameter and low sample size. The average impact parameter of 2DAs appears to be at least 5 times less than the average impact parameter of Mg II absorbers at around 7 kpc, as compared to the Mg II average impact parameter of 48 kpc. This indicates that 2DAs are likely associated with the disk components of high redshift galaxies. Our simulations show that an imaging survey with better than 1.4 arcsec is necessary to possibly detect host galaxies of 2DAs. 

Due to the low amount of residue apparent, the stacking method may not prove to be effective for low impact parameter host galaxy systems under seeing conditions of 1.4 arcsec or worse. The normalization of the PSF most likely subtracts off the host galaxy along with the QSO. However, we also demonstrated that at low impact parameters, the stacking technique is still capable of drawing upper limits for impact parameters. 

Our implementation of the Principal Component Analysis machine learning algorithm for fitting astronomical source profiles was an improvement over single-source PSF approximation, thereby reducing the noise level in the PSF frame by $5 - 6$ times. To our best knowledge, this is the first time that the PCA method was used in PSF subtraction for studying quasar host galaxies. PCA is extremely flexible in terms of approximation power and noise level containment, making it a strong method of fitting PSF profiles. In particular, despite the differences in methodology, our SED (Figure 10) produced is very similar to Zibetti et al.'s (2007). This opens up new possibilities for approximating stellar source profiles to much higher accuracies in future studies, while decreasing the overall noise level in the PSF frame during subtraction. In addition, our flood-fill masking algorithm was much more adaptive in effectively masking non-circular sources, thereby reducing the amount of data being unnecessarily masked by up to 3 times depending on the source's geometry.

For the 2DAs, hopefully the recent release of SDSS Data Release 14 will yield more samples, allowing us to bin the absorbers by redshift and REW of the 2175 \AA\ absorption line and perform more comprehensive studies in the future.

\section*{Acknowledgements}

We thank the anonymous referee for his/her insightful feedback that greatly improved the accuracy and robustness of our procedure and data analysis. We also thank the Sloan Digital Sky Survey for providing open access to imaging data. BZ acknowledges the University of Florida and its Department of Astronomy for supporting him during the summer of 2017. We would also like to thank Dr. Stefano Zibetti for his assistance in developing the stacking method, as well as Dr. Denis Burgarella for his guidance in using CIGALE.

Funding for the SDSS and SDSS-II has been provided by the Alfred P. Sloan Foundation, the Participating Institutions, the National Science Foundation, the U.S. Department of Energy, the National Aeronautics and Space Administration, the Japanese Monbukagakusho, the Max Planck Society, and the Higher Education Funding Council for England. The SDSS Web Site is http://www.sdss.org/.

Funding for SDSS-III has been provided by the Alfred P. Sloan Foundation, the Participating Institutions, the National Science Foundation, and the U.S. Department of Energy Office of Science. The SDSS-III web site is http://www.sdss3.org/.

The SDSS is managed by the Astrophysical Research Consortium for the Participating Institutions. The Participating Institutions are the American Museum of Natural History, Astrophysical Institute Potsdam, University of Basel, University of Cambridge, Case Western Reserve University, University of Chicago, Drexel University, Fermilab, the Institute for Advanced Study, the Japan Participation Group, Johns Hopkins University, the Joint Institute for Nuclear Astrophysics, the Kavli Institute for Particle Astrophysics and Cosmology, the Korean Scientist Group, the Chinese Academy of Sciences (LAMOST), Los Alamos National Laboratory, the Max-Planck-Institute for Astronomy (MPIA), the Max-Planck-Institute for Astrophysics (MPA), New Mexico State University, Ohio State University, University of Pittsburgh, University of Portsmouth, Princeton University, the United States Naval Observatory, and the University of Washington.

SDSS-III is managed by the Astrophysical Research Consortium for the Participating Institutions of the SDSS-III Collaboration including the University of Arizona, the Brazilian Participation Group, Brookhaven National Laboratory, Carnegie Mellon University, University of Florida, the French Participation Group, the German Participation Group, Harvard University, the Instituto de Astrofisica de Canarias, the Michigan State/Notre Dame/JINA Participation Group, Johns Hopkins University, Lawrence Berkeley National Laboratory, Max Planck Institute for Astrophysics, Max Planck Institute for Extraterrestrial Physics, New Mexico State University, New York University, Ohio State University, Pennsylvania State University, University of Portsmouth, Princeton University, the Spanish Participation Group, University of Tokyo, University of Utah, Vanderbilt University, University of Virginia, University of Washington, and Yale University.

This research has made use of the NASA/IPAC Extragalactic Database (NED) that is operated by the Jet Propulsion Laboratory, California Institute of Technology, under contract with the National Aeronautics and Space Administration.







\bsp	
\label{lastpage}
\end{document}